\documentstyle[11pt]{article}

\makeatletter
\@addtoreset{equation}{section}
\makeatother

\makeatletter
\def\citen#1{%
\edef\@tempa{\@ignspaftercomma,#1, \@end, }
\edef\@tempa{\expandafter\@ignendcommas\@tempa\@end}%
\if@filesw \immediate \write \@auxout {\string \citation {\@tempa}}\fi
\@tempcntb\m@ne \let\@h@ld\relax \def\@citea{}%
\@for \@citeb:=\@tempa\do {\@cmpresscites}%
\@h@ld}
\def\@ignspaftercomma#1, {\ifx\@end#1\@empty\else
   #1,\expandafter\@ignspaftercomma\fi}
\def\@ignendcommas,#1,\@end{#1}
\def\@cmpresscites{%
 \expandafter\let \expandafter\@B@citeB \csname b@\@citeb \endcsname
 \ifx\@B@citeB\relax 
    \@h@ld\@citea\@tempcntb\m@ne{\bf ?}%
    \@warning {Citation `\@citeb ' on page \thepage \space undefined}%
 \else
    \@tempcnta\@tempcntb \advance\@tempcnta\@ne
    \setbox\z@\hbox\bgroup 
    \ifnum0<0\@B@citeB \relax
       \egroup \@tempcntb\@B@citeB \relax
       \else \egroup \@tempcntb\m@ne \fi
    \ifnum\@tempcnta=\@tempcntb 
       \ifx\@h@ld\relax 
          \edef \@h@ld{\@citea\@B@citeB }%
       \else 
          \edef\@h@ld{\hbox{--}\penalty\@highpenalty
            \@B@citeB }%
       \fi
    \else   
       \@h@ld\@citea\@B@citeB
       \let\@h@ld\relax
 \fi\fi%
 \def\@citea{,\penalty\@highpenalty\hskip.13em plus.1em minus.1em}%
}
\def\@citex[#1]#2{\@cite{\citen{#2}}{#1}}%
\def\@cite#1#2{\leavevmode\unskip
  \ifnum\lastpenalty=\z@\penalty\@highpenalty\fi
  \ [{\multiply\@highpenalty 3 #1
      \if@tempswa,\penalty\@highpenalty\ #2\fi 
    }]\spacefactor\@m}
\makeatother

\let\a=\alpha \let\b=\beta  \let\d=\delta 
   \let\i=\iota \let\k=\kappa
    \let\p=\pi 
     
      \let\G=\Gamma   
  \let\S=\Sigma  \let\F=\Phi

\def\nn{\nonumber} \def\bd{\begin{document}} \def\ed{\end{document}}
\def\ds{\documentstyle} \let\fr=\frac \let\bl=\bigl \let\br=\bigr
\let\Br=\Bigr \let\Bl=\Bigl 
\let\bm=\bibitem
\let\na=\nabla
\let\pa=\partial \let\ov=\overline 
\newcommand{\be}{\begin{equation}} 
\newcommand{\ee}{\end{equation}} 
\def\ba{\begin{array}}
\def\ea{\end{array}}
\def\ft#1#2{{\textstyle{{\scriptstyle #1}\over {\scriptstyle #2}}}}
\def\fft#1#2{{#1 \over #2}}
\def\del{\partial}
\def\vp{\varphi}
\def\sst#1{{\scriptscriptstyle #1}}
\def\oneone{\rlap 1\mkern4mu{\rm l}}
\def\simequiv{\buildrel\sim\over=}
\def\td{\tilde}
\def\wtd{\widetilde}
\def\ie{{\it i.e.\ }}
\def\im{{\rm i}}
\def\dalemb#1#2{{\vbox{\hrule height .#2pt
        \hbox{\vrule width.#2pt height#1pt \kern#1pt
                \vrule width.#2pt}
        \hrule height.#2pt}}}
\def\square{\mathord{\dalemb{6.8}{7}\hbox{\hskip1pt}}}
\def\R{\rlap{\rm I}\mkern3mu{\rm R}}
\def\E{\rlap{\rm I}\mkern3mu{\rm E}}
\def\Z{\rlap{\sf Z}\mkern3mu{\sf Z}}
\def\cA{{\cal A}\!\!\!\!\!{\cal A}}
\def\F#1#2{{ F_{#1}^{(#2)} }}
\def\cF#1#2{{ {\cal F}_{#1}^{(#2)} }}
\def\sw{{\sst W}}
\def\st{{\sst T}}
\def\bog{Bogomol'nyi\ }
\newcommand{\ho}[1]{$\, ^{#1}$}
\newcommand{\hoch}[1]{$\, ^{#1}$}
\newcommand{\bea}{\begin{eqnarray}} 
\newcommand{\eea}{\end{eqnarray}} 
\newcommand{\ra}{\rightarrow}
\newcommand{\lra}{\longrightarrow}
\newcommand{\Lra}{\Leftrightarrow}
\newcommand{\ap}{\alpha^\prime}
\newcommand{\bp}{\tilde \beta^\prime}
\newcommand{\tr}{{\rm tr} }
\newcommand{\Tr}{{\rm Tr} } 
\newcommand{\NP}{Nucl. Phys. }

\newcommand{\cern}{\it TH Division, CERN, CH-1211 Geneva 23, Switzerland}
\newcommand{\sissatamphys}{\it SISSA, Via Beirut No. 2-4, 34013 Trieste, 
Italy and\\
Center for Theoretical Physics,
Texas A\&M University, College Station, Texas 77843}
\newcommand{\ens}{\it Laboratoire de Physique Th\'eorique de l'\'Ecole
Normale Sup\'erieure\hoch{4}\\
24 Rue Lhomond - 75231 Paris CEDEX 05, France}
\newcommand{\ic}{\it The Blackett Laboratory, Imperial College,\\
Prince Consort Road, London SW7 2BZ, UK}

\newcommand{\auth}{M.S. Bremer\hoch{\star\,1,2},  
H. L\"u\hoch{\dagger}, C.N. Pope\hoch{\ddagger\,3} and
K.S. Stelle\hoch{\natural\star\,1}}

\begin{document}
\begin{flushright}
\hfill{CERN-TH/97-271}\\
\hfill{CTP TAMU-40/97}\\
\hfill{Imperial/TP/96-97/72}\\
\hfill{LPTENS-97/42}\\
\hfill{SISSARef.\ 122/97/EP}\\
\hfill{hep-th/9710244}\\
\hfill{Oct. 1997}\\
\end{flushright}

\begin{center}
{\bf\large Dirac Quantisation Conditions and Kaluza-Klein 
Reduction}

\vspace{15pt}
\auth

\vspace{10pt}

{\hoch{\star}\ic}

\vspace{5pt}
{\hoch{\dagger}\ens}

\vspace{5pt}
{\hoch{\ddagger}\sissatamphys}

\vspace{5pt}
{\hoch{\natural}\cern}

\vspace{10pt}

\underline{ABSTRACT}
\end{center}

        We present the form of the Dirac quantisation condition for
the $p$-form charges carried by $p$-brane solutions of supergravity
theories. This condition agrees precisely with the conditions obtained
in lower dimensions, as is necessary for consistency with Kaluza-klein
dimensional reduction. These considerations also determine the charge
lattice of BPS soliton states, which proves to be a universal
modulus-independent lattice when the charges are defined to be the
canonical charges corresponding to the quantum supergravity symmetry
groups.

{\vfill\leftline{}\vfill
\footnoterule\vskip  5pt
{\footnotesize \hoch{1} Research supported in part by the
European Commission under TMR contract \vskip -12pt} \vskip 10pt
{\footnotesize \hoch{\phantom{1}}
     ERBFMRX-CT96-0045. \vskip -12pt} \vskip 14pt
{\footnotesize \hoch{2} Research supported in part by the
European Commission under TMR contract \vskip -12pt} \vskip 10pt
{\footnotesize \hoch{\phantom{2}}
     ERBFMBI-CT97-2344. \vskip -12pt} \vskip 14pt
{\footnotesize  \hoch{3} Research supported in part by DOE 
Grant DE-FG03-95ER40917 and the \vskip -12pt} \vskip 10pt
{\footnotesize \hoch{\phantom{3}} EC Human Capital and Mobility 
Programme under contract ERBCHBGT920176.\vskip  -12pt} \vskip 14pt
{\footnotesize
        \hoch{4} Unit\'e Propre du Centre National de la Recherche
Scientifique, associ\'ee \`a l'\'Ecole Normale Sup\'erieure
\vskip -12pt} \vskip 10pt
{\footnotesize \hoch{\phantom{4}} et \`a l'Universit\'e de Paris-Sud
\vskip -12pt} \vskip 10pt}

\pagebreak
\setcounter{page}{1}

\section{Introduction}

     The global ({\it i.e.}\ rigid) supersymmetry algebra of $D=11$
supergravity takes the form \cite{hvp,azcarragaetal}
\bea
\{ Q , Q \} &=& (C\G ^{\underline{M}}) P_{\underline{M}} +{1 \over 2} 
(C\G ^{\underline{M}_1 \underline{M}_2}) Z_{\underline{M}_1 
\underline{M}_2} \nonumber \\
                 &+& {1 \over 5!} (C\G ^{\underline{M}_1 \underline{M}_2 
\dots \underline{M}_5}) Y_{\underline{M}_1 \underline{M}_2 \cdots 
\underline{M}_5}, \label{eq0}
\eea
where $\underline{M} = (0, M)$ are tangent-space $D=11$ indices, 
$C$ is the charge conjugation matrix, $P_{\underline{M}}$ is the $D=11$ 
ADM momentum and $Z_{\underline{M} _1 \underline{M} _2}$ and 
$Y_{\underline{M} _1 \underline{M} _2 \cdots \underline{M} _5}$ are 
the analogues of the `central charges' of the $D=4$ supersymmetry algebra.
These charges are not `central,' in $D=11$, evidently since they carry
non-trivial Lorentz indices. Upon Kaluza-Klein dimensional reduction to $D=4$,
these indices become labels for the various Lorentz-scalar central charges of
the descendant $N=8$, $D=4$ supergravity theory. The occurrence of such
tensorial charges is one of the striking features of the $D=11$ theory, and is
fundamental for the subject of $p$-branes, which are the carriers of such
charges.

     In this paper, we shall study the implications of the Dirac quantisation
condition for tensorial charges such as $Z_{\underline{M} _1 \underline{M}
_2}$ and $Y_{\underline{M} _1 \underline{M} _2 \cdots \underline{M} _5}$, both
in their original higher-dimensional incarnations and also in regard to
the quantisation conditions on their  dimensionally-reduced descendants. Along
the way, we shall cast further light on the structure of the $p$-brane charge
lattice, showing the way in which charge-unit scales are set.

The paper is organised as follows: in section \ref{sec:charges}, we
shall discuss with some care the construction of the tensorial
charges, focusing in particular on the topological class of curves in
the $p$-brane transverse space that leads to ostensibly scalar charges
nonetheless being labeled by $p$-forms. In section \ref{dirac}, we
shall use this information to derive the Dirac quantisation condition
for $p$-branes, following \cite{nepomechie} and \cite{teit}, but
emphasising the existence of `Dirac-insensitive' configurations. In
section \ref{dimred}, we shall show how this charge-quantisation
picture accords perfectly with the quantisation conditions obtained in
lower dimensions by dimensional reduction. In section \ref{wavesnuts}
we shall extend this picture to include quantisation of wave and NUT
solutions by classical arguments which nonetheless fit in neatly with
the Dirac condition. This is then extended in section \ref{dyons} and
section \ref{intersections} to cover dyonic and self-dual $p$-branes
and the relation between Dirac insensitive configurations and
intersecting $p$-branes. This leads us in section \ref{web} to a
discussion of the web of relations between charge scales for all
$p$-branes in various dimensions. This discussion will be similar to
that presented in Refs \cite{sch} and \cite{alw}, but we shall make
the point that by considering the implications of T duality together with
some special `scale-setting' $p$-brane species, namely the self-dual
3-branes in $D=10$ type IIB theory and the dual pairs of D0 branes and
$(D-4)$ branes, charge scales in $D\le 10$ supergravity theories are
in fact determined without recourse to M theory. (In fact the
relations that we derive can be interpreted as supporting evidence for
the M-theory conjecture.)  The resulting charge lattices will then be
discussed in section \ref{chargelattice}. In the Appendix, we discuss
the structure of Dirac quantisation conditions for dyons in even
dimensions, which have the familiar antisymmetric structure in $D=4k$
dimensions, but become symmetric in $D=4k+2$ dimensions such as for
the self-dual cases of strings in $D=6$ and 3-branes in $D=10$.

\section{$p$-form charges}
\label{sec:charges}

A charged $p$-brane embedded in a $D$-dimensional supergravity
background naturally carries a conserved $(p+1)$-form
current\footnote{Such a current may be considered to exist even for
non-singular $p$-brane solutions to the supergravity field
equations. Even though the singularity structure of the solution does
not then necessitate the introduction of source terms, it is
nevertheless possible to couple a $p$-brane source consistently to a
non- singular $p$-brane supergravity background.  For further detail
on source placement in supergravity solutions, cf.\ \cite{bremer}.}
$J_{p+1}$. For a $p$-brane carrying electric (magnetic) charge $Q_e$
($Q_m$) under some $n$-form field strength $F_n$, where $n=p+2$
($n=D-p-2$), this current appears as a source term on the RHS of the
field equation (Bianchi identity) for the field strength. The
conservation condition for the current can be concisely formulated as
$d*J_{p+1}=0$. It implies that the charge \be Q_\S = \int
_\S{*J_{p+1}},
\label{eq1}
\ee 
where $\S$ is a $(D-p-1)$-dimensional spacelike surface, is conserved
in time, provided that the current flowing through the boundary $\pa
\S$ vanishes.  Note that unless $p=0$, the integration surface $\S$ is
a subsurface of the chosen spacelike hypersurface that serves as the
integration domain for ordinary scalar charge integrals.\footnote{In
  this paper, we shall mostly consider static solutions for which a
  natural `rest frame' set of coordinates exists, thus defining a
  preferred spacelike hypersurface as the general arena for charge
  integrals such as (\ref{eq1}).} Thus, for $p\ne 0$, the integration
surface $\S$ is not unique, i.e. there is no unique embedding of $\S$
into this spacelike hypersurface. The $p$-brane charge $Q_\S$ may thus
in general be expected to depend on the choice of integration surface
$\S$. We shall see in the following, however, that this dependence is
essentially topological.

     Consider accordingly now the dependence of the $p$-brane charge on 
the choice of the integration surface $\S$. In a coordinate system 
$x^{\underline{M}}=\{t,x^M\}$ $(M=1,2,\ldots (D-1))$ the integral (\ref{eq1})
can be written as
\be
Q_\S={1\over p!}\int_\S J^{0M_1M_2\cdots M_p}\, d\S_{M_1M_2\cdots M_p},
\label{eq2}
\ee
where $d\S_{M_1M_2\cdots M_p}$ is the coordinate volume element on $\S$ 
and the $p$-brane current density $J^{0 M_1 M_2 \cdots M_p}$ is given 
by an integral over the $p$-brane worldvolume ${\cal W} _{p+1}$
\be
J(x) ^{0 M_1 M_2 \cdots M_p}= Q_{\rm e/m} \, \int _{{\cal W}_{p+1}}
{\d (t-T)\, \d (x-X) \ dT \wedge dX^{M_1} \wedge dX^{M_2} \wedge 
\cdots \wedge dX^{M_p}}.
\label{eq3}
\ee 
Here, $X^{\underline{M}}=(T, X^M)$ are the coordinates of the
$p$-brane and $Q_{\rm e/m}$ is the electric or magnetic source charge.
Owing to the presence of the $\d$-functions in the integral
(\ref{eq3}), the non-zero contribution to $Q_\S$ comes from the
intersection of $\S$ with the worldvolume ${\cal W}_{p+1}$ of the
$p$-brane \cite{hennteit}. The dimension of this intersection is zero
and hence $\S\, \cap \, {\cal W}_{p+1}$ consists in general of a
finite number of discrete points. Each point in the intersection
contributes $\pm Q_{\rm e/m}$ to the integral, according to the
orientation with which ${\cal W}_{p+1}$ pierces $\S$, {\it i.e.}\ from
`above' or from `below'. It is also possible for the integration
surface $\S$ to be tangential to the $p$-brane worldvolume ${\cal
W}_{p+1}$ at the intersection point, or indeed for $\S$ and ${\cal
W}_{p+1}$ to overlap partially (in which case there are an infinite
number of intersection points). However, in such cases an
infinitesimal deformation of $\S$ near the intersection point would
result in no intersection at all, or in pairs of intersections, with
one from `above' and one from `below' in each pair. The net
contribution to the corresponding charge $Q_\S$ is always zero in such
cases. Hence, one may effectively ignore both `tangential' intersection
points and `overlapping' integration surfaces. The two possible values
of $Q_\S$ are thus $\{Q_{\rm e/m}, 0\}$, according to whether the
intersection consists of an odd or even number of points. These two
values for $Q_\S$ naturally partition the set of all integration
surfaces into two distinct subsets corresponding to the two values of
the $p$-brane charge. We shall next show that these subsets can also
be thought of as topological equivalence classes of integration
surfaces.

To this end, consider deforming the integration surface $\S$
infinitesimally to a nearby surface $\S^\prime$. Mathematically, this
is achieved by considering the flow $\Phi_V$ induced by a spacelike
vector field $V$ normal to $\S$.  The flow $\Phi_V$ maps a point
${\cal P} \in \S$ to a nearby point ${\cal P}^\prime$, which is
obtained by going an infinitesimal parameter distance along the
integral curve (through $p$) of the vector field $V$.\footnote{The
  integral curve of $V$ through a point $\cal P$ with coordinates
  $x^M({\cal P})$ is given by the solution to the differential
  equation $dx^M/d\sigma = V^M(x(\sigma))$, with initial condition
  $x^M(0)=x^M({\cal P})$.}  Mapping each point ${\cal P} \in \S$ to a
nearby point ${\cal P}^\prime$, we get the deformed surface
$\S^\prime$.  The infinitesimal change in the charge $Q_\S$ is then
given by the integral (over $\S$) of the Lie derivative along $V$ of
$*J_{p+1}$.  Because the $p$-brane current is conserved, ${\cal
  L}_V*J_{p+1}=d\,\i_V*J_{p+1}$, where $\i_V$ denotes the interior
product with the vector field $V$.  Using Stokes' theorem, we may
write the change in $Q_\S$ as an integral over the boundary of $\S$:
\be \d Q_\S=\int_{\pa\S}\i_V*J_{p+1} ={1\over
  p!}\int_{\pa\S}J^{0M_1M_2\cdots M_p}V^N\, d\S_{M_1M_2\cdots M_pN}.
\label{eq4}
\ee
Because the $p$-brane current density $J^{0M_1M_2\cdots M_p}$ is
non-zero only on the $p$-brane worldvolume ${\cal W}_{p+1}$, $\d Q_\S$
thus vanishes unless $\pa\S$ intersects the $p$-brane.  Since the
charge integrals are always carried out within a fixed spacelike
hypersurface in spacetime ({\it i.e.}\ at fixed `time'), the question
of whether $\pa\S$ intersects the $p$-brane at a fixed time is more
precisely the question whether $\pa\S$ intersects the intersection of
${\cal W}_{p+1}$ with the chosen spacelike hypersurface. The vanishing
of $\d Q_\S$ implies that two integration surfaces $\S_1$ and $\S_2$
give rise to the same $p$-brane charge if their {\em boundaries} can
be continuously deformed into one another {\em without intersecting
  the $p$-brane} in the course of the deformation. The set of
integration surfaces is therefore naturally partitioned into
equivalence classes of surfaces whose boundaries can be continuously
deformed into one another without crossing the $p$-brane. All surfaces
belonging to a given equivalence class give rise to the same value of
the $p$-form charge.  If the spacetime manifold is simply connected,
the converse of this statement also holds. Any two integration
surfaces $\S_1$ and $\S_2$ such that $Q_{\S_1}=Q_{\S_2}$ must belong
to the same equivalence class, {\em i.e.}\ their boundaries can be
continuously deformed into one another without intersecting the
$p$-brane worldvolume.\footnote{Note the equality of the charges
  associated with the integration surfaces $\S_i$ implies that the
  numbers of intersections of the $\S_i$ themselves with the $p$-brane
  differ (at most) by an even number, where this even difference of
  contributions to the charges cancels out in positive/negative pairs.
  In such a case, however, the integration surface boundaries
  $\partial \S_i$ may still be deformed into one another without
  intersecting the $p$-brane.}  We therefore conclude that the set of
equivalence classes of integration surfaces for a given $p$-brane
solution consists of only two points, which are naturally associated
with the two values $\{Q_{\rm e/m}, 0\}$ of the $p$-brane charge.

     What remains to be done now is to find a representative member of the
equivalence class of integration surfaces which give rise to a
non-zero $p$-brane charge. One might think it natural for the
$p$-brane charge to be labeled by this representative integration
surface. However, it is clear from the above that the spatial
$p$-brane section of the $p$-brane worldvolume determines the set of
topologically equivalent integration surfaces that give rise to a
non-zero charge. Hence, we expect that the form structure of the
$p$-brane charge is ultimately characterised by the configuration of
this spatial section itself. To make these ideas more precise, let us
concentrate on $p$-branes of infinite spatial extent with $\R^p$
topology. In fact, it will turn out that these, together with
$p$-branes wrapped around compact dimensions, are the only ones that
can carry non-zero $p$-form charge \cite{hennteit}. Now, in order for
a $p$-brane of infinite spatial extent to have a finite energy
density, it must be asymptotically flat. `Asymptotically flat' is here
taken to consist of the following two conditions: (i) the deviation of
the metric from the flat Minkowski metric vanishes `sufficiently
rapidly' as one approaches transverse spatial infinity {\em in
  spacetime} (for more details, see \cite{ght}), and (ii) the
transverse vibrations of the $p$-brane worldvolume vanish
`sufficiently rapidly' as one approaches transverse spatial infinity
{\em on the worldvolume}. The second condition in particular implies
that the tangent vectors to the $p$-brane worldvolume at spatial
infinity on the worldvolume must point along asymptotically flat
directions in spacetime.  These directions then define the asymptotic
orientation of the $p$-brane.  For sake of definiteness, we assume
that the spatial section of the $p$-brane worldvolume is
asymptotically oriented along the $\{x^1,x^2,\ldots x^p\}$ directions,
where $x^{\underline{M}}= \{t,x^i,y^m\}$ ($i=1,2,\ldots p$ and
$m=(p+1),\ldots (D-1)$) is an asymptotically flat coordinate system.
The remaining asymptotically flat spatial directions $\{y^m\}$ define
the asymptotic transverse space, which we denote by $\S ^{1 2 \cdots
  p}$.

We now note that the asymptotic transverse space $\S^{1 2 \cdots p}$
of a given $p$-brane solution is a $(D-p-1)$-dimensional spacelike
surface which necessarily intersects a spatial section of the
$p$-brane worldvolume in an {\em odd} number of points. It is
therefore a representative integration surface for the equivalence
class of integration surfaces that give rise to a non-zero $p$-brane
charge. As this asymptotic transverse space is entirely determined by
the asymptotic orientation of the spatial section of the $p$-brane
worldvolume, it is natural to {\em label} the $p$-brane charge by the
asymptotic orientation of this spatial section. Another way of seeing
why this labeling is natural is to realise that the boundaries of all
integration surfaces that give rise to a non-zero $p$-brane charge
must have the same topology, because they can be continuously deformed
into one another. The boundary of the asymptotic transverse space is
topologically equivalent to the $(D-p-2)$-dimensional sphere
$S^{D-p-2}$ that totally surrounds the spatial section of the
$p$-brane worldvolume. So, the equivalence class of integration
surfaces that give rise to a non-zero $p$-brane charge can be
characterised by the fact that the boundary of each surface in the
equivalence class is topologically equivalent to a $S^{D-p-2}$ that
totally surrounds the $p$-brane.  This topological condition on the
boundaries of the integration surfaces is manifestly dependent on the
asymptotic orientation of the $p$-brane worldvolume's spatial section.

It now is clear why only infinite $p$-branes or $p$-branes that are
wrapped around compact dimensions can carry non-zero $p$-form charges
\cite{hennteit}.\footnote{While it is true that only infinite or
  wrapped $p$-branes can carry the $p$-form charges appearing in the
  supersymmetry algebra (\ref{eq12}), conserved charges defined by
  (\ref{eq1}) with surfaces $\Sigma$ that do not necessarily extend
  out to transverse infinity also have important r\^oles to play in
  the theory.  For example, the conservation of charges of this sort,
  with $\partial\Sigma$ closely looping around a $p$-brane segment,
  may be used to derive the intersection rules between different
  $p$-branes \cite{surgery}. There does not seem to be any way to
  associate such `small $\partial\Sigma$' charges with $p$-forms,
  however.} Only such $p$-branes can be `captured' by a bounding
$S^{D-p-2}$ surface at infinity, recalling that continuous
deformations of such a bounding surface can be made provided they do
not intersect the $p$-brane. Thus, for a $p$-brane of finite extent,
any candidate integration surface could have its boundary deformed and
completely moved away from the $p$-brane at infinity, in which case
the resulting charge is clearly zero.  Another way of saying this is
to note that integration volumes for $p$-form charge integrals will be
intersected an even number of times by $p$-branes of finite extent,
with positive and negative contributions to the charge canceling out.

We may encode the information about the value of the $p$-brane charge
and the asymptotic orientation of an infinite $p$-brane's spatial
worldvolume section by defining a $p$-{\em form} charge $Q_p$ whose
magnitude $|Q_p|$ is equal to the electric or magnetic charge $Q_{\rm
  e/m}$ of the $p$-brane and which is proportional to the $p$-brane's
asymptotic spatial section volume form. For example, if the $p$-brane
is asymptotically oriented as above, $Q_p=Q_{\rm e/m}dx^1 \wedge dx^2
\wedge \cdots \wedge dx^p$. Note that is always possible to rotate the
asymptotically flat coordinate basis in such a way that the $p$-form
charge $Q_p$ is proportional to a single $p$-form basis element, which
labels the asymptotic orientation of the $p$-brane.
 
     We conclude this section by giving some examples of $p$-form charges in 
$D=11$ supergravity and by recalling the r\^oles they play in determining 
the residual supersymmetries of a $p$-brane configuration. It follows from the 
field equation (Bianchi identity) for the $4$-from field strength 
$F_4$ of $D=11$ supergravity in the presence of electrically (magnetically) 
charged $2$-brane ($5$-brane) sources that the canonical 
$2$-brane electric charge $Q_\S$ ($5$-form magnetic charge $P_{\tilde{\S}}$ ) 
are given in terms of $F_4$ by
\bea
Q_\S &=& \int _{\pa \S} (*F_4 -\ft12 A_3\wedge F_4)\label{eq9} \\
P_{\tilde{\S}} &=& \int _{\pa \tilde{\S}} F_4. 
\label{eq10}
\eea
Note that the canonical charges (\ref{eq9}) and (\ref{eq10}) differ
from the source charges in (\ref{eq4}) by a factor of $\k ^2 _{11}$,
{\it i.e.}\ $Q_{\rm canonical}= \k ^2 _{11} Q_{\rm source}$. This is
because the $11$-dimensional gravitational coupling constant $\k
_{11}$ ($\kappa_{11}^2=8\pi G$) only multiplies the $D=11$
supergravity action (as $1\over 2\k_{11}^2$) but not the $p$-brane
  source action. Henceforth, all charges will be understood to be
  canonical, unless otherwise stated. Both the source and the
  canonical charges here are dimensionful; we shall later discuss
  charge lattices in terms of dimensionless charges in sections
  \ref{web} and \ref{chargelattice}.

It is easy to verify, for the standard charge $Q_{\rm e}$ $2$-brane
solution \cite{ds} oriented along the $\{ x^1 x^2\}$ directions, that
$Q_\S=0$ unless $\pa\S$ is topologically equivalent to the
7-dimensional sphere at transverse spatial infinity surrounding the
$2$-brane. This implies that the only non-vanishing independent
component of the $2$-form electric charge $Q_2$ is $Q_{1 2}= Q_{\rm
  e}$. Similarly, for the charge $Q_{\rm m}$ $5$-brane \cite{guven}
oriented along the $\{ x^1 x^2 \cdots x^5\}$ directions, one must have
that $\pa\tilde{\S}$ is the $4$-dimensional sphere at transverse
spatial infinity surrounding the $5$-brane. We therefore get
$P_{12345}=Q_{\rm m}$ with all other independent components of $P_5$ equal to
zero. This is of course exactly what was expected from the general
analysis given above.

The relevance of the electric $2$-form charge $Q_2$ and the magnetic
$5$-form charge $P_5$ for determining the supersymmetries of a
$p$-brane configuration in $D=11$ supergravity comes from their
appearance in the `maximally extended' $D=11$ supersymmetry algebra
(\ref{eq0}) The algebra (\ref{eq0}) can be derived using the Nester
form of $D=11$ supergravity \cite{azcarragaetal} whence it becomes
apparent that the {\em spatial} components of $Z_2$ and $Y_5$ are
proportional to the components of $Q_2$ and $P_5$, the explicit
relations being $Z_2=\kappa_{11}^{-\fft{14}9}Q_2$ and
$Y_5=\kappa_{11}^{-\fft39}P_5$, the factors of $\kappa_{11}$ being
chosen to correct for dimensionality. The mixed time-space components
of $Z_2$ and $Y_5$ are associated with charges in Kaluza-Klein vacua
\cite{st}. Ignoring these and setting also the spatial momentum $P_M$
to zero, we can rewrite (\ref{eq0}) in the Majorana representation,
where $C=\G ^0$, as
\be
\{ Q , Q \} = {\cal M} +{1 \over 2} \G ^{0 M_1 M_2} Z_{M_1 M_2} 
+ {1 \over 5!} \G ^{0 M_1 M_2 \cdots M_5} Y_{M_1 M_2 \cdots M_5}, 
\label{eq12}
\ee
where ${\cal M}$ is the ADM mass of the $p$-brane. This form of the 
$D=11$ supersymmetry algebra is appropriate for a static $p$-brane 
configuration. It makes explicit the dependence of the algebra, and hence 
the  supersymmetries of the $p$-brane, on the $2$- and $5$-form charges $Q_2$ 
and $P_5$. In the next section, we shall formulate the Dirac 
quantisation condition for $Q_2$ and $P_5$.

\section{Dirac quantisation conditions for $p$-branes}
\label{dirac}

The usual Dirac quantisation condition between electric and magnetic
charges in four dimensions admits a straightforward generalisation to
extended objects in higher dimensions. Specifically, if an
electrically charged $p$-brane exists in the presence of its
magnetically charged dual $(\tilde{p}=D-p-4)$-brane, then by
considering the phase of the wavefunction of the $p$-brane as it is
transported around the $\tilde{p}$-brane, one can derive a
quantisation condition\footnote{See also \cite{dkl} and references
therein for more detailed discussions of Dirac quantisation conditions
in string theory.} between the electric and magnetic charges
\cite{nepomechie,teit}, namely $Q_{\rm e} \, Q_{\rm m} = {\rm integer}
\times 2\p\kappa^2$. Strictly speaking, this is a quantisation
condition for the {\em magnitudes} of the electric and magnetic
charges only. However, as we have seen in the previous section, a $(p
\ne 0)$-brane charge carries additional indices labeling the
asymptotic directions along which the spatial section of the $p$-brane
worldvolume is oriented. The magnitude of the charge together with the
asymptotic spatial orientation of the $p$-brane are encoded in the
$p$-form charge $Q_p$. It is always possible to find an asymptotically
flat coordinate basis in which $Q_p$ is proportional to a single basis
$p$-form.  This $p$-form points along the asymptotically flat spatial
directions of the $p$-brane worldvolume. The magnitude of this
$p$-form charge is equal to the electric or magnetic charge carried by
the $p$-brane, {\it i.e.}\ $|Q_p|= Q_{\rm e/m}$. The Dirac
quantisation condition can then be rewritten in terms of the electric
$p$-form charge $Q_p$ and the magnetic $\tilde{p}$-form charge
$P_{\tilde{p}}$ as $|Q_p| \, |P_{\tilde{p}}| = {\rm integer} \times
2\p\kappa^2$. Here, we want to generalise this condition by taking the
asymptotic directions of the worldvolumes into account. To be precise,
we shall show that there is a quantisation condition involving the
magnitudes of $Q_p$ and $P_{\tilde{p}}$ only if there is no overlap
between the asymptotic orientations of the spatial sections of the
electric $p$-brane and the magnetic $\tilde{p}$-brane worldvolumes.

To begin, let us recall how one arrives at the Dirac quantisation
condition for $p$-branes \cite{nepomechie, teit}. We take as an
example the quantisation of the electric $2$-brane charge and the
magnetic $5$-brane charge in $D=11$.  Suppose we bring a $2$-brane
`probe' with charge $|Q_2|=Q_{\rm e}$ into a $5$-brane $D=11$
supergravity background with charge $|P_5|=Q_{\rm m}$.  The $2$-brane
and the $5$-brane share a common time direction, but for the moment we
shall not make any assumptions about the relative orientation of their spatial
worldvolume sections. We denote the $2$-brane coordinates by
$X^{\underline{M}}=(T, X^M)$ $(M=1, 2, \ldots 10)$; $A_3$ is the
singular $3$-form potential for the $4$-form field strength $F_4$ in
the presence of the $5$-brane. Now consider deforming the $2$-brane ({\it
i.e.}\ a spatial section of the 2-brane worldvolume) through
a spacelike path at a constant time, with identical initial and final
$2$-brane configurations. The chosen spatial section of the $2$-brane
worldvolume traces out a closed spacelike three-dimensional surface
$\cal W$.  Of course, the surface $\cal W$ does not correspond to a
physical motion of the $2$-brane, but nevertheless it needs to be taken
into account in a quantum-mechanical description ({\it e.g.}\ in a
path integral formulation) of the $2$-brane. As the $2$-brane is taken
around $\cal W$, the $2$-brane's wavefunction acquires a phase factor
\be
{\rm exp}\left({{\rm i}Q_{\rm e} \over 3! \k ^2} \oint _{\cal W} 
A_{M_1 M_2 M_3}\, dX^{M_1} \wedge \, dX^{M_2} \wedge \, dX^{M_3}\right).
\label{eq13}
\ee 
Here, the charge multiplying the integral is a canonically defined charge 
and hence we need the additional $\k ^{-2}$ factor ({\it c.f.}\ 
section~\ref{sec:charges}). 

Using Stokes' theorem, we can rewrite the integral 
of $A_3$ over $\cal W$ as the integral of $F_4=dA_3$ over any 
`capping' surface $\cal M$ whose boundary is $\cal W$, 
{\it i.e.}\ $\pa {\cal M}={\cal W}$:
\bea
&&{Q_{\rm e} \over 3! \k ^2} \oint
_{\cal W}{ A_{M_1 M_2 M_3}\,  dX^{M_1} \wedge \, dX^{M_2} \wedge \,
dX^{M_3}} \nonumber \\ &&\qquad=\ {Q_{\rm e} \over 4! \k ^2} \int _{\cal
M} F_{M_1 M_2 M_3 M_4}\,  dX^{M_1} \wedge \, dX^{M_2} \wedge \, dX^{M_3}
\wedge \, dX^{M_4} \nonumber \\ &&\qquad=\ {Q_{\rm e} \over \k ^2}
{\Phi}_{\cal M}, \label{eq14}
\eea
where ${\Phi}_{\cal M}$ is the flux of $F_4$ through the cap 
$\cal M$. Taking two different choices ${\cal M}_1$ and 
${\cal M}_2$ for the cap, one can end up with a discrepancy 
$\Phi_{{\cal M}_1}-\Phi_{{\cal M}_2}$ if the two caps taken together 
form a  four-dimensional closed surface 
${\cal M}_{\rm total}={\cal M}_1\cup{\cal M}_2$ that captures the net 
flux from the 5-brane, which by Gauss' law is equal to the magnetic 
charge, {\it i.e.}\ $\Phi_{\rm total}=Q_{\rm m}$. This may be viewed
as the  discrepancy between zero and $\int_{{\cal M}_{\rm total}}dA$,
which would have vanished if the gauge potential $A_3$ for the
$5$-brane background had been everywhere non-singular. In the 
quantum-mechanical description, this discrepancy gives rise to a phase factor
\be
{\rm exp} \left(\im {Q_{\rm e}Q_{\rm m} \over \k ^2}\right) \label{eq15}
\ee
in the wavefunction of the 2-brane probe. The Dirac quantisation condition 
arises from the requirement that this phase factor equal unity, {\it i.e.}\
that
\be 
Q_{\rm e}Q_{\rm m} = 2\pi\kappa^2 n\ ,\qquad n\in\Z\ .\label{eq16} 
\ee

     The above derivation produces a Dirac quantisation condition involving 
the {\em magnitudes} of the 2-brane's electric $2$-form charge $Q_2$ and 
the 5-brane's magnetic $5$-form charge $P_5$ whenever the orientations of 
the 2-brane and the 5-brane allow one to construct a surface 
${\cal M}_{\rm total}$, with $\cal W$ as the boundary separating 
${\cal M}_1$ and ${\cal M}_2$, that captures the 5-brane's total magnetic 
flux $\Phi_{\rm total}$. 

     Let us now take the asymptotic orientations of 
the $2$-brane and $5$-brane into account to assess exactly under what 
conditions one can construct ${\cal M}_{\rm total}$. We can always choose 
an asymptotically flat coordinate system $x^{\underline{M}}=(t, x^M)$ 
such that the spatial section of the $5$-brane worldvolume lies
asymptotically along the $\{ x^1, x^2, \ldots x^5 \}$ directions. It
follows from the  general considerations in section~\ref{sec:charges} that the
non-vanishing  component of the magnetic $5$-form charge $P_5$ is
\be
P_{1 2 3 4 5} = \int _{\pa \tilde{\S}} F_4 = Q_{\rm m}, \label{eq17}
\ee
where $\pa \tilde{\S}$ is topologically equivalent to the $S^4$ at transverse
spatial infinity that surrounds the $5$-brane. This identifies the closed 
surface that captures the total $5$-brane flux as 
${\cal M}_{\rm total} = S^4$. The two capping surfaces ${\cal M}_1$ 
and ${\cal M}_2$ for the $2$-brane path $\cal W$ must therefore 
correspond to the `northern' and `southern' hemispheres of the $S^4$, 
with $\cal W$ being the `equatorial' $S^3$. Note that 
${\cal M}_{\rm total}$ lies entirely within the asymptotic transverse 
space of the $5$-brane.  
 
Let's now bring the $2$-brane into play, which we assume to be
asymptotically oriented along the $\{ x^{M_1}, x^{M_2}\}$ directions,
and see how we can generate a closed spacelike surface $\cal W$, by
deforming the 2-brane configuration of fixed 2-form charge through a
closed path $\cal W$ which is topologically equivalent to an
`equatorial' $S^3$ at transverse spatial infinity.\footnote{There are
  two ways to view the path $\cal W$. One is as the motion of a
  2-brane held rigidly in the same shape as it is taken around in the
  5-brane background, but compactifying the points at the 2-brane's
  spatial infinity, so that $\R^2\times S^1\rightarrow S^3$. Another
  way to view $\cal W$ is to keep the asymptotic configuration of the
  2-brane fixed, but to deform its shape, expanding out a `bubble' of
  2-brane that then sweeps out an $S^3$ path surrounding the magnetic
  charge centre. The latter appears to be the point of view adopted in
  \cite{nepomechie}.} The asymptotic orientation of the 2-brane is
part of its boundary conditions, and hence has to be maintained
throughout the motion around the closed path.

Two distinctly different cases now arise, according to whether or not
the asymptotic orientations of the spatial sections of the $2$-brane
and $5$-brane worldvolumes partially coincide. Suppose first that they
do coincide, {\it i.e.}\ that $M_1 \in \{1, 2, 3, 4, 5 \}$ or $M_2 \in
\{1, 2, 3, 4, 5 \}$. Recall that this means that, as we approach
spatial infinity on the $2$-brane's worldvolume, one (or both) of the
tangent vectors to the $2$-brane spatial worldvolume section starts
pointing in directions along which the $5$-brane is asymptotically
oriented.  It follows that {\em any} three-dimensional surface $\cal
W$ generated by taking the given $2$-brane around a closed path shares
the following property: there exists a subspace of $\cal W$ with
tangent vectors lying parallel to spatial infinity on the $2$-brane
worldvolume, for which one (or two) of these tangent vectors point
along the directions of the asymptotic orientation of the $5$-brane.
Compare this to the `equatorial' $S^3$ which we wish to generate. Its
tangent vectors are everywhere linearly independent from those of the
spatial worldvolume section of the $5$-brane. To see this, it is
useful to note that the `equatorial' $S^3$ lies at transverse spatial
infinity in the 5-brane spacetime, where the metric is {\em flat}, by
the usual boundary conditions for the $5$-brane $D=11$ supergravity
background. Alternatively, the geometry of the spacelike hypersurfaces
of the $5$-brane background at transverse spatial infinity has
topology $\R^5 \times S^4$, where $\R^5$ corresponds to the flat
spatial worldvolume of the $5$-brane and $S^4$ to the four-sphere at
infinity.  Clearly, the `equatorial' $S^3$ lies within the $S^4$
factor, which is everywhere transverse to the spatial worldvolume of
the $5$-brane. Thus, in order for the capping surface ${\cal M}_{\rm
  total}$ to capture the 5-brane's flux, the path $\cal W$ must lie
within the topological equivalence class of paths `surrounding' the
5-brane as discussed in section \ref{sec:charges}, with tangent
vectors everywhere independent of those of the 5-brane worldvolume.
Any other path can be deformed into one for which the flux captured is
zero.  It follows that one cannot establish a quantisation condition
for the electric $2$-brane and magnetic $5$-brane charges whenever the
spatial worldvolumes of the $2$-brane and $5$-branes have an
asymptotic coincidence of orientations. Such a $2$-brane/$5$-brane
configuration will be called Dirac-insensitive.

Whenever there is no asymptotic coincidence of orientations between
the spatial sections of the $2$-brane and the $5$-brane worldvolumes,
{\it i.e.}\ when $M_1 \notin \{1, 2, 3, 4, 5 \}$ and $M_2 \notin \{1,
2, 3, 4, 5 \}$, then the asymptotic directions along which the
$2$-brane is oriented will also lie in the asymptotic transverse space
of the $5$-brane.  There will then be no topological considerations
preventing the construction of ${\cal M}_{\rm total}$ as a union of
capping surfaces of $\cal W$. The above argument leading to the
quantisation of the electric $2$-brane and magnetic $5$-brane charges
then applies, and thus one obtains the quantisation condition
(\ref{eq16}) involving the magnitudes of the electric $2$-form charge
$Q_2$ and the magnetic $5$-form charge $P_5$.

     One can verify the orientation dependence of the Dirac quantisation
condition explicitly for a case in which the spatial worldvolumes of both the
$2$-brane  and the $5$-brane are strictly flat. A flat $5$-brane oriented along
the $\{x^1, x^2, \ldots x^5 \}$ directions with charge $Q_{\rm m}$ is 
given by the following classical solution \cite{guven}
\bea 
ds_{11}^2 &=&
H^{-\ft13}\,dx^\mu dx^\nu \eta_{\mu\nu} +
H^{\ft23}\,dy^mdy^m\nonumber\\
F_4 &=& *(d^6x\wedge dH^{-1}), \label{eq18}
\eea
where $H$ is a harmonic function of the $\{ y^m \}$ with a pole strength such 
that $P_{1 2 3 4 5}=Q_{\rm m}$. Furthermore, let the flat $2$-brane `probe' 
be oriented along the $\{x^{M_1}, x^{M_2}\}$ directions and construct a 
closed spacelike surface $\cal W$ as described above. The quantum mechanical
phase factor (\ref{eq13}) associated with $\cal W$ now  reduces to
\be
{\rm exp}\left({{\rm i}Q_{\rm e} \over \k ^2} \oint _{\cal W} 
{\partial X^P\over\partial\sigma} \,A_{M_1 M_2 P} \right),
\label{eq19}
\ee 
where the ${\partial X^P\over\partial\sigma}$ vector points in the
`third' direction along the path $\cal W$, {\it i.e.}\ the one not
lying along the spatial section of the $2$-brane worldvolume; the
coordinate volume element on $\cal W$ has also been suppressed. Now
note that since the non-vanishing components of $F_4$ in
(\ref{eq18}) all point in directions transverse to the $5$-brane, one
can always find a gauge in which the potential $A_3$ is entirely
transverse as well. The phase factor (\ref{eq19}) is then trivially equal
to unity whenever $M_1$ or $M_2$ point along the spatial worldvolume
of the $5$-brane, because the integrand vanishes identically in such
cases. This confirms that whenever the (strictly flat) $2$-brane and
$5$-brane overlap, one cannot establish a quantisation condition for
their charges.

Let us now summarise the general situation for the $D=11$ Dirac
quantisation condition in $D=11$ between the electric $2$-form charge
$Q_2$ and the magnetic $5$-form charge $P_5$. The requirement that
there be no coincidence between the asymptotic orientations of the spatial
worldvolume sections of the $2$-brane and the $5$-brane is formulated 
concisely as the condition that $Q_2\wedge P_5\ne 0$. In this case,
the magnitudes $|Q_2|=Q_{\rm e}$ and $|P_5|=Q_{\rm m}$ obey a
quantisation condition. Therefore, the Dirac quantisation condition,
taking into account the asymptotic orientations of the $2$-brane and
the $5$-brane, may be written in general
\be
Q_2 \wedge P_5 = 2\pi\kappa^2 n {Q_2 \wedge P_5 \over |Q_2|\,|P_5|}\ ,
\qquad n\in \Z\ .
\label{eq20}
\ee
Equivalently, one may express this condition as
\be
(|Q_2|\,|P_5|- 2\pi\kappa^2 n)Q_2 \wedge P_5 =0\ .\label{eqxxxx}
\ee

     Note that (\ref{eq20}) or (\ref{eqxxxx}) becomes vacuous whenever 
the $2$-brane and the $5$-brane asymptotically align, {\it i.e.}\ whenever
$Q_2\wedge P_5=0$. Such cases are precisely the Dirac-insensitive
configurations. However, these configurations clearly form a subset of
measure zero within the general configuration space of a 2-brane and a
5-brane. Since a small rotation is all that is necessary
to change an  insensitive configuration into one for which Dirac quantisation
becomes  applicable, generic 2-branes and 5-branes in $D=11$ must all satisfy
the Dirac condition. 

     It is nonetheless worth noting the existence of the Dirac
insensitive configurations for several reasons. One is that this will be
relevant for the comparison that we shall shortly make between higher and
lower dimensional Dirac quantisation conditions in theories that are related by
dimensional reduction.  Another reason for taking note of the insensitive
configurations is the observation that the insensitive configurations coincide
with those  of intersecting $p$-branes, which also possess zero-force
properties  related to the preservation of residual supersymmetry. 
Yet a third reason concerns the sharpness with which the $p$-form charges are
defined in quantum mechanics.

One might consider that quantum fluctuations would smear out the
orientation of a $p$-brane, so that the measure-zero set of
Dirac-insensitive configurations might disappear at the quantum level.
We shall not enter into a detailed discussion of the question, but
shall be content to make some indicative observations as to why this
insensitive set may nonetheless persist at the quantum level. We have
seen that the conserved charges carried by $p$-branes are naturally
$p$-form objects, and these carry sharply-defined information about
the object's asymptotic spatial orientation. Note that these $p$-form
charges would have as conjugate variables the time-independent modes
of the $p$-form gauge parameters $\Lambda_p$ for the $(p+1)$-form
gauge potentials $A_{p+1}$ ($\delta A_{p+1}=d\Lambda_p$). Since these
gauge parameters do not contain physical degrees of freedom, one might
expect there to be no inconsistency with having the $p$-form charges
sharply defined at the quantum level. Another way to think of this is to
recall that in order for a $p$-brane to carry a non-vanishing charge,
it must either be infinite in extent or must be wrapped around a
compact dimension, with the charge arising for essentially topological
reasons whenever it is possible to `capture' the $p$-brane with the
boundary of the charge integration surface. Infinite $p$-branes have
sharply-defined asymptotic orientations because their moments of
inertia are infinite. Moreover, $p$-branes wrapped around compact
dimensions have sharply-defined orientations when they are in their
ground states, while their excited states have energies that tend to
infinity as one shrinks the radius of the compact dimension.
Accordingly, the Dirac-insensitive configurations may have a more
persistent role than they might at first seem to have, even though
they constitute only a subset of measure zero within the set of all
$p$-brane configurations.

\section{Dimensional reduction and Dirac quantisation conditions}
\label{dimred}

    Now let us consider the Dirac quantisation condition for
$p$-branes in the context of dimensional reduction to $D=4$. In this
case, one has only to deal with electrically and magnetically charged
particles (\ie black holes).  For this, one first needs to study how the
charges of the higher-dimensional $p$-branes are related to those of their
dimensionally reduced descendants.  

     The $D$-dimensional bosonic Lagrangian
resulting from the dimensional reduction of eleven-dimensional
supergravity takes the form
\bea
{\cal L} &=& eR -\ft12 e\, (\del\vec\phi)^2 -\ft1{48}e\, e^{\vec a\cdot
\vec\phi}\, F_4^2 -\ft{1}{12} e\sum_i
e^{\vec a_i\cdot \vec\phi}\, (F_3^{(i)})^2
-\ft14 e\, \sum_{i<j} e^{\vec a_{ij}\cdot \vec\phi}\, (F_2^{(ij)})^2
\label{dgenlag}\\
&& -\ft14e\, \sum_i e^{\vec b_i\cdot \vec\phi}\, ({\cal F}_2^{(i)})^2
-\ft12 e\, \sum_{i<j<k} e^{\vec a_{ijk} \cdot\vec \phi}\,
(F_1^{(ijk)})^2 -\ft12e\, \sum_{i<j} e^{\vec b_{ij}\cdot \vec\phi}\,
({\cal F}_1^{(ij)})^2 + {\cal L}_{\sst{FFA}}\ ,\nn
\eea
where the `dilaton vectors' $\vec a$, $\vec a_i$, $\vec a_{ij}$,
$\vec a_{ijk}$,
$\vec b_i$, $\vec b_{ij}$ are constants that characterise the couplings of
the dilatonic scalars $\vec \phi$ to the various gauge fields.
They are given by \cite{lpsol}
\vfill\eject
\bea
&&F_{\sst{MNPQ}}\qquad\qquad\qquad\qquad\qquad\qquad\qquad\qquad
{\rm vielbein}\nonumber\\
{\rm 4-form:}&&\vec a = -\vec g\ ,\nonumber\\
{\rm 3-forms:}&&\vec a_i = \vec f_i -\vec g \ ,\nonumber\\
{\rm 2-forms:}&& \vec a_{ij} = \vec f_i + \vec f_j - \vec g\ ,
\qquad\qquad\qquad\qquad\qquad \,\,\, \,\vec b_i = -\vec f_i \ ,
\label{dilatonvec}\\
{\rm 1-forms:}&&\vec a_{ijk} = \vec f_i + \vec f_j + \vec f_k -\vec g
\ ,\qquad\qquad\qquad\qquad\vec b_{ij} = -\vec f_i + \vec f_j\ ,\nonumber
\nonumber
\eea
where 
\be
\vec g \cdot \vec g = \ft{2(11-D)}{D-2}, \qquad
\vec g \cdot \vec f_i = \ft{6}{D-2}\ ,\qquad
\vec f_i \cdot \vec f_j = 2\delta_{ij} + \ft2{D-2}\ .\label{gfdot}
\ee
The field strengths are associated with the gauge potentials in the
obvious way; for example $F_4$ is the field strength for $A_3$, $F_3^{(i)}$
is the field strength for $A_2^{(i)}$, {\it etc}.  The complete expressions
for the Kaluza-Klein modifications to the various field strengths are
given in \cite{lpsol}, as are the cubic Wess-Zumino terms ${\cal L}_{FFA}$
coming from the $F_4\wedge F_4\wedge A_3$ term in the eleven-dimensional
Lagrangian. The eleven-dimensional and $D$-dimensional 
metrics are related by \cite{lpsol,cjlp}
\be
ds_{11}^2 = e^{\ft13 \vec g\cdot\vec\phi} \, ds_{\sst D}^2 +
\sum_i e^{2\vec\gamma_i\cdot\vec\phi}\, (h^i)^2\ ,\label{met}
\ee
where $\vec \gamma_i=\ft16\vec g -\ft12\vec f_i$, and 
\be
h^i=dz^i + {\cal A}_1^i + {\cal A}_0^i{}_j\, dz^j\ .
\ee

     We shall define the electric and magnetic charges for each field
strength to be the canonical Noether charges
\be
Q_{\rm e} = \int (e^{\vec c\cdot\vec\phi} \, *F + K(A))\ ,\qquad 
Q_{\rm m} = \int \tilde F \ ,\label{charges}
\ee
where $F=\tilde F + \cdots$ is the field strength, with the ellipses
representing the Kaluza-Klein modifications, $\tilde F=dA$, and $\vec
c$ is the dilaton vector corresponding to $F$, as given in
(\ref{dilatonvec}).  The term $K(A)$ represents the contributions
coming from the Wess-Zumino terms ${\cal L}_{FFA}$ in the
$D$-dimensional Lagrangian.  The vector $\vec c$ denotes the dilaton
vector for $F$, as given above.  Let us now see how these charges are
related to charges in $D=11$.  We begin by considering the cases where
the $D$-dimensional field strengths come from the dimensional
reduction of $\hat F_4$ in $D=11$. From this, the following fields can
arise in $D$ dimensions: $F_4$, $\F3i$, $\F2{ij}$ or $\F1{ijk}$.  The
expansion of the eleven-dimensional 4-form field strength $\hat F_4 $
in terms of the $D$-dimensional fields is as follows \cite{lpsol}:
\be
\hat F_4 = F_4 + \F3i\wedge h^i + \ft12 \F2{ij}\wedge h^i\wedge h^j +
\ft16 \F1{ijk}\wedge h^i\wedge h^j\wedge h^k\ .
\ee
It is easy to show from (\ref{met}) that the eleven-dimensional Hodge 
dual $\hat *$ of $\hat F_4$ is related to the $D$-dimensional Hodge duals 
$*$ of the $D$-dimensional fields by
\be
\hat * \hat F_4 = e^{\vec a\cdot\vec\phi}\, *F_4 \wedge v +
 e^{\vec a_i\cdot\vec\phi}\, *\F3i \wedge v^i +
\ft12 e^{\vec a_{ij}\cdot\vec\phi}\, *\F2{ij} \wedge v^{ij}
+\ft16 e^{\vec a_{ijk}\cdot\vec\phi}\, *\F1{ijk} \wedge v^{ijk}\ ,
\ee
where we have defined
\bea
&&v=\ft1{(11-D)!} \epsilon_{i_1\cdots i_{\sst{11-D}}}\, h^{i_1}\wedge \cdots
\wedge h^{i_{\sst{11-D}}} \ ,\quad 
v_i=\ft1{(10-D)!} \epsilon_{ii_2\cdots i_{\sst{11-D}}}\, h^{i_2} \wedge\cdots
\wedge h^{i_{\sst{11-D}}} \ ,\\
&&v_{ij}=\ft1{(9-D)!} \epsilon_{iji_3\cdots i_{\sst{11-D}}}\, 
h^{i_3}\wedge \cdots
\wedge h^{i_{\sst{11-D}}} \ ,\quad 
v_{ijk}=\ft1{(8-D)!} \epsilon_{ijki_4\cdots i_{\sst{11-D}}}\, 
h^{i_4} \wedge\cdots
\wedge h^{i_{\sst{11-D}}} \ .\nn\label{vdef}
\eea From (\ref{charges}), one may then obtain the expressions for the 
eleven-dimensional charges in terms of the $D$-dimensional ones.  The
results are given in Table 1 below.

\bigskip\bigskip

\centerline{
\begin{tabular}{|c|c|c|c|c|}\hline
& $F_4$ & $\F3i$ & $\F2{ij}$ & $\F1{ijk}$ \\ \hline\hline
Electric $Q^{11}_{\rm e}=$ & $Q^{\sst D}_{\rm e}\, V$ & $Q^{\sst
  D}_{\rm e}\, 
\fft{V}{L_i}$ & $Q^{\sst D}_{\rm e}\, \fft{V}{L_i\, L_j}$ & 
$Q^{\sst D}_{\rm e}\, \fft{V}{L_i\, L_j\, L_k}$\\ \hline
Magnetic $Q^{11}_{\rm m}=$ & $Q^{\sst D}_{\rm m}$ & $Q^{\sst D}_{\rm m}\, 
{L_i}$
& $Q^{\sst D}_{\rm m}\, {L_i\, L_j}$ & $Q^{\sst D}_{\rm m}\, {L_i\,
L_j\, L_k}$\\ \hline
\end{tabular}}
\bigskip

\centerline{Table 1: Relations between $Q^{11}$ and $Q^{\sst D}$}
\bigskip

In Table 1, $L_i$ denotes the period of the compactification
coordinate $z^i$, and the compactification volume is $V= \int d^{11-D}
\vec z = \prod_{i=1}^{11-D} L_i$.  Note that the expressions for the
eleven-dimensional charges in terms of the $D$-dimensional
canonically-defined charges do not depend on the scalar moduli of the
$D$-dimensional theory.

    From the results in Table 1, we see that any Dirac quantisation
condition between a conjugate pair of electric and magnetic charges in
$D$ dimensions, namely
\be
Q_{\rm e}^{\sst D}\, Q_{\rm m}^{\sst D} = 2\pi\, \kappa_{\sst D}^2\, n\ ,
\label{ddirac}
\ee
agrees precisely with the Dirac quantisation condition 
\be
Q^{11}_{\rm e}\, Q^{11}_{\rm m} = 2\pi \, \kappa_{11}^2 \, n\label{11dirac}
\ee
between the membrane and the 5-brane charges in $D=11$, since the
gravitational couplings in the two cases are clearly related by
\be
\kappa_{11}^2 = V\, \kappa_{\sst D}^2\ .\label{kappa}
\ee
The result (\ref{11dirac}) gives the quantisation condition on the
magnitudes of the membrane and 5-brane charges.  As we have discussed
in the previous section, the true quantisation condition must take
account of the form structure of the charges for the extended objects.
The form indices correspond to the worldvolume spatial-section indices
of the objects. Let us consider the special case where one reduces to
$D=4$, so that the conjugate pairs of objects subject to Dirac
quantisation conditions are simply electric and magnetic black holes,
for which the charges carry no indices.  Then, from the Dirac
quantisation condition for any electric and magnetic charge carried by
one given 2-form field strength in $D=4$, we can deduce the corresponding
quantisation condition on membranes and 5-branes in $D=11$. In
particular, a black hole supported by an electric charge for the field
strength $\F2{ij}$ in $D=4$ will oxidise to a membrane in $D=11$ with
spatial worldvolume coordinates $z^i$ and $z^j$. Conversely, a
magnetic black hole supported by the same field strength will oxidise
to a 5-brane with world-volume directions complementary to these, \ie
$\{z^{k_1},\ldots, z^{k_5}\}$, where $i,j,{k_1},\ldots, {k_5}$ are all
different \cite{classp}. Thus we obtain a Dirac quantisation condition
involving the membrane and 5-brane charges when the worldvolume spatial
sections of the two objects share no common directions. This agrees
precisely with the results obtained in the previous section.

In order for a Dirac quantisation condition in a lower dimension $D$,
as expressed in (\ref{ddirac}), to be inherited from the original
$D=11$ condition (\ref{11dirac}), one requires at each step of
dimensional reduction a mixed combination of a diagonal dimensional
reduction for one $p$-brane and a vertical dimensional reduction for
the other. Otherwise, the electric and magnetic brane solutions in the
lower dimension will not be supported by the same field strength, and
so no Dirac quantisation condition will arise. If one persists
nonetheless in making other combinations of dimensional reduction, the
orientation-sensitivity of the Dirac condition (\ref{eq20}) will give
a nil result again, fitting in precisely with the expected pattern of
quantisation conditions in lower dimensions.

For example, consider making a diagonal dimensional reduction for both
a 2-brane and a 5-brane in $D=11$. This will produce in $D=10$ a
1-brane ({\it i.e.}\ a string) supported by a 3-form field strength
and a 4-brane supported by a 4-form field strength, so there should be
no Dirac quantisation condition.  But taking into account the
orientation sensitivity expressed in (\ref{eq20}), one sees that this
is precisely what happens, because in order to make a diagonal
dimensional reduction for both of the $D=11$ branes, they must share a
spatial worldvolume direction, in which case (\ref{eq20}) gives zero,
as required.

As another example, consider making vertical dimensional reductions
for a 2-brane and a 5-brane. Vertical dimensional reduction works by
first making a `stack' of branes in the compactification direction
\cite{vertical}, in order to generate a translational symmetry in a
direction transverse to the worldvolume.  Both electric and magnetic
charges now need to be interpreted as charge densities per unit length
in this compactification direction, in addition to the natural
$p$-brane interpretation of charge as a density per unit $p$ volume.
Letting the period in the compact direction be $L$, the total phase
factor for a closed-circuit deformation as discussed in section
\ref{dirac} will be $e^{\im\kappa_{11}^{-2} Q_eQ_mL^2}$, with one
factor of $L$ in the exponent coming from each stack of branes. In the
compactification limit, one has
$\lim_{L\rightarrow0}{L^2\over\kappa_{11}^2}=0$, so no Dirac
quantisation condition is inherited in the lower dimension. This case
should be contrasted with that of a mixed diagonal/vertical reduction,
where only one of the branes is stacked up and with the corresponding
charge being given a density interpretation in the compactification
dimension. For such a mixed reduction configuration, one has a phase
factor $e^{\im\kappa_{11}^{-2} Q_eQ_mL}$, and, noting the relation
(\ref{kappa}) between the gravitational couplings, one obtains
precisely the lower-dimensional Dirac quantisation condition
(\ref{ddirac}). Note also that the relative orientations of the two
branes in the higher dimension do not in this case correspond to a
Dirac-insensitive configuration.

Thus we have seen that there is a complete accord between the form
structure of the Dirac quantisation condition for $p$-branes
(\ref{eq20}) and the natural Dirac quantisation conditions occurring in
lower dimensions, obtained from the higher-dimensional one by
Kaluza-Klein dimensional reduction. The absence of Dirac conditions
between descendant branes that are not duals of each other in the
lower dimension does not mean, however, that there are no relations at
all between the spectra of such branes. We shall see in sections
\ref{web} and \ref{chargelattice} that there are `incidental'
relations between non-dual branes, which are implied by T duality
together with the existence of certain `scale-setting' $p$-branes.

\section{Waves, NUTs and quantisation conditions}
\label{wavesnuts}

In the previous section, we showed how the Dirac quantisation rules
for charges constructed using components of the $D=11$ 4-form field
strength are related under dimensional reduction.  One should also
consider the Dirac quantisation rules for charges constructed using
the Kaluza-Klein vectors occurring in lower dimensions, and should
investigate how these rules are related to the quantisation rules in
$D=11$.  However, although objects carrying these charges in the lower
dimension will be ordinary $p$-branes, {\it e.g.}\ D0-branes in
$D=10$, their oxidations to $D=11$ will give either pp waves, in the
case of electric charges, or Taub-NUT-like monopoles ({\it i.e.}\ 
NUTs), in the case of magnetic charges.

     Let us begin by considering the reduction from $D=11$ to $D=10$.
The metrics in the two dimensions are related by
\be
ds_{11}^2 = e^{\ft16\phi} \, ds_{10}^2 + e^{-\ft43\phi}\,
(dz+{\cal A}_1)^2\ .\label{met1110}
\ee
The total canonically normalised momentum $P_z$ carried by a wave
moving in the $z$
direction in eleven dimensions is given by \cite{adm}
\be
P_z = \int \hat * d \hat e^z\ ,\label{11mom}
\ee
where the integral is restricted to the $11-2=9$ dimensional boundary of
the spacelike hypersurface in $D=11$ spacetime, 
$\hat e^z=e^{-\ft23\phi}(dz+{\cal A}_1)$ is the vielbein for the
eleventh direction, and $\hat *$ denotes the eleven-dimensional
Hodge dual, as given earlier.  Note that (\ref{11mom}) defines the total
momentum of a plane wave with translational invariance in the
direction of propagation, which therefore would diverge in an
uncompactified spacetime. At transverse spatial infinity, the dilaton
$\phi$ tends to the constant value $\phi_0$, and so we find that the momentum
(\ref{11mom}) can be rewritten in a fashion ready for compactification as
\be
P_z = e^{-\ft56\phi_0}\, \int * {\cal F}_2\wedge(dz+{\cal A}_1) \ ,
\ee
where $*$ denotes the ten-dimensional Hodge dual. Since the electric charge
in $D=10$ is given by $Q_{\rm e} = \int e^{-\ft32\phi}\, *{\cal F}_2$,
it follows that the total momentum of the pp wave in $D=11$ is related to
the electric charge of the Kaluza-Klein vector in $D=10$ by
\be
P_z = Q_{\rm e}\, e^{\ft23\phi_0}\, L\ ,
\ee
where $L$ is the period of the compactifying coordinate $z$.  From
(\ref{met1110}) we see that the physical radius of the eleventh
dimension is given by $R_{11}= e^{-\ft23\phi_0}\, L/(2\pi)$, and so we have
\be
P_z = \fft{Q_{\rm e}\, L^2}{2\pi R_{11}} \ .\label{wavecharge}
\ee 
At the quantum level, the momentum of a wave must be quantised so that
an integer number of wavelengths fit around the circle.  Thus we must
have that
\be
P_z = \fft{n\, \kappa_{11}^2}{R_{11}}\ ,\label{wavequant}
\ee
where the $\kappa_{11}^2$ factor compensates for the dimension $(\rm length)^8$
of the canonically normalised momentum (compare the relation between canonical
and source charges given in section \ref{sec:charges}). Putting this together
with (\ref{wavecharge}), we see that the electric charge in
$D=10$ carried by the D0-brane obtained from the pp wave by
dimensional reduction is quantised according to
\be
Q^{10}_{\rm e}= \fft{2\pi n\, \kappa_{11}^2}{L^2} = \fft{2\pi n\, 
\kappa_{10}^2}{L}
\ .\label{d0quant}
\ee

Note that the quantisation of the D0-brane charge implied by these
eleven-dimensional considerations is quite different from the usual
kind of Dirac quantisation condition, in that it occurs without
needing the existence of a conjugate magnetically-charged object. If,
nevertheless, one does consider the magnetic charge at the same time,
it becomes important to verify that the above quantisation is
consistent with the Dirac quantisation condition. The magnetic dual of
the D0-brane in $D=10$ is a D6-brane.  From the eleven-dimensional
point of view, it is a NUT.  The associated `NUT charge' $Q_{\sst{\rm
NUT}}=\int {\cal F}_2$ is classically quantised, in the sense that
in order for the eleven-dimensional metric (\ref{met1110}) to be
non-singular, the period of $z$ must satisfy $L= 1/k\, \int {\cal F}_2
=Q_{\sst{\rm NUT}}/k$, where $k$ is any integer.  Since the magnetic
charge carried by the D6-brane in $D=10$ is also given by $Q_{\rm m}=
\int{\cal F}_2 =Q_{\sst{\rm NUT}}$, it follows that we have a
classical discretisation of the allowed magnetic charge values
\be
Q^{10}_{\rm m} = k\, L\ , \qquad k\in\Z\ .\label{d6quant}
\ee
Thus, combining this with (\ref{d0quant}), we see indeed that one
obtains a result that is consistent with the $D=10$ Dirac quantisation
condition, namely
\be
Q^{10}_{\rm e}\, Q^{10}_{\rm m} = 2\pi\, \kappa_{10}^2\, q\ ,
\ee
where $q$ is an integer.

An important distinction between the quantisations (\ref{d0quant}) and
(\ref{d6quant}) for D0-branes and D6-branes in $D=10$, as compared to
the standard Dirac quantisation rules, is that an absolute scale is
obtained for the individual electric and magnetic charges of the
D0-branes and D6-branes. Usually, in a Dirac quantisation condition,
there is an arbitrariness under which the unit of electric charge can
be scaled up by any factor, with a simultaneous corresponding scaling
down of the unit of magnetic charge. As we shall see in section
\ref{web}, the absolute scale of the Kaluza-Klein pp wave quantisation
will play an important r\^ole in giving all of the charges in type IIA
theory the same scale, and this gives an important consistency condition
for the conjectured existence of M-theory.

\section{Dirac quantisation for dyonic and self-dual $p$-branes}
\label{dyons}
 
    Dyonic $p$-branes can arise in any even dimension, and self-dual
$p$-branes can arise in $D=4n+2$ dimensions (in the case of Lorentzian
spacetime signature). In fact, it is appropriate to divide the
discussion of all dyonic $p$-branes into two classes corresponding
to $D=4n+2$ and to $D=4n$.  We shall begin by considering dyons in
$D=4n$.  First, we recall that in $D=4$, the electric and magnetic
charges $e$ and $g$ carried by two dyons must satisfy the Dirac
quantisation condition
\be
e_1\, g_2 -e_2 \, g_1 = 2\pi\, n\ .
\ee
An analogous result can also be obtained for dyonic membranes in
$D=8$.  These membranes can be obtained from simple single-charge
electric or magnetic membranes by making a transformation under the
$SL(2,\R)$ factor of the $SL(3,\R)\times SL(2,\R)$ global symmetry
group of $D=8$ maximal supergravity \cite{dyonic}. The Dirac quantisation
rule for dyonic membranes can be derived simply by acting with
an $SL(2,\R)$ transformation on the standard $Q_{\rm e}\, Q_{\rm m}=
2\pi\, \kappa_8^2\,  n$ quantisation rule for a pure electric and a
pure magnetic membrane.  The result, which is $SL(2,\R)$ invariant, is
\be
Q_1^T\, \Omega\, Q_2 = 2\pi\,\kappa_8^2\,  n\ ,\label{d8dyon}
\ee
where $Q=(Q_{\rm e}, Q_{\rm m})$, and $\Omega$ is the
$SL(2,\R)$-invariant antisymmetric matrix
\be
\Omega= \pmatrix{0 & 1\cr -1 & 0}\ .
\ee
To be precise, as we have explained in section 2, the charges in this case
really carry indices, corresponding to the world-volume directions for
the two dyons.  The quantisation condition (\ref{d8dyon}) arises only
in cases where the two individual dyonic membranes have non-aligned 
orientations of their spatial world-volumes, so that the wedge
product of their charge forms is non-zero.  It is worth recalling
that, in both $D=4$ and $D=8$, the quantisation conditions become
vacuous for the case of two dyons whose electric and magnetic charges
are in the same ratio.  One way in which this may be seen is by
observing that a global symmetry transformation can be used to rotate both
of the dyons simultaneously to a purely electric or purely magnetic
form, for which there is no quantisation condition.

     The situation is quite different in $D=4n+2$ dimensions.  Let us
first consider the $D=6$ case. There are five 3-forms $\F3i$ in $D=6$
which, together with their duals, form a ten-dimensional irreducible
vector representation under the global symmetry group\footnote{See,
for example, \cite{cjlp} for a detailed discussion of this symmetry. A
general discussion of duality symmetries in dimensions $4k$ and
$4k+2$, including cases with matter coupling, may be found in
\cite{adf}.} $O(5,5)$ .  The $O(5,5)$-invariant Dirac quantisation
condition for dyonic strings with five electric charges $\vec Q$ and
five magnetic charges $\vec P$ will then be
\be
\ft12 \pmatrix{\vec Q_1+\vec P_1\ , & \vec Q_1-\vec P_1} \, \pmatrix{\oneone &
0 \cr 0  & -\oneone}\, \pmatrix{\vec Q_2+\vec P_2 \cr \vec Q_2-\vec P_2}
= 2\pi\,\kappa_6^2\, n\ ,\label{sympdyon}
\ee
and so
\be
\vec Q_1\cdot\vec P_2 + \vec Q_2\cdot \vec P_1 =  2\pi\,\kappa_6^2\,
n\ .\label{d6dyon}
\ee
If we specialise to the case where only one of the five field
strengths carries the electric and magnetic charges, this result
reduces to 
\be
Q_{\rm e}^{(1)}\, Q_{\rm m}^{(2)} + Q_{\rm e}^{(2)}\, Q_{\rm m}^{(1)}
= 2\pi\,\kappa_6^2\, n\ ,\label{dyond6}
\ee
where $Q_{\rm e}^{(1)}$ and $Q_{\rm m}^{(1)}$ are the electric and
magnetic charges of the first dyon, and $Q_{\rm e}^{(2)}$ and 
$Q_{\rm m}^{(2)}$ are the charges for the second dyon. 

This {\em symmetric} quantisation condition may seem surprising, but
one may also verify that the symmetric structure is correct for
$(2k+1)$-forms in dimensions $D=4k+2$ by generalising the original
quantum phase arguments of Ref.\ \cite{schwinger}, noting that in
$D=4k+2$ dimensions, one has (for Minkowski signature) $\wtd{\wtd
  F}_{2k+1}= F_{2k+1}$. In such cases, the phase acquired by taking
one dyon with charges $(Q_e,Q_m)$ around another with (electric,
magnetic) sources $(J_{2k},\wtd J_{2k})$ is $\kappa^{-2}(Q_e\int\ast
J_{2k} + Q_m\int\ast\wtd J_{2k})$, confirming the symmetric
structure.\footnote{Although not widely commented upon in the
  literature, this symmetric structure has been recognised in a
  number of contexts, including in particular the basic case of $D=2$,
 and in the $O(5,5)$-invariant six-dimensional supergravity
  \cite{julia}. A clear treatment of the dyonic quantisation condition
  in $D=4k+2$ dimensions based upon the requirement that Dirac strings
  be unobservable has been given in \cite{dght}. This argument fixes
  the overall coefficients in the conditions
  (\ref{sympdyon}--\ref{dyond6}). We shall see in Section \ref{web}
  that this coefficient is also essential for consistency with the
  ordinary Dirac condition obtained in one less dimension by dimensional
  reduction from (\ref{sympdyon}--\ref{dyond6}).}

As a consequence of this symmetric structure, the quantisation
condition for the six-dimensional dyonic strings survives even when
both of the dyons have the same ratio of electric to magnetic
charges.\footnote{Naively, one might think that dyons of equal
  charge-ratio could be rotated to purely electric or purely magnetic
  complexions, just as can be done in $4n$ dimensions.  However this
  is not possible here, since there is no duality symmetry between
  electric and magnetic strings coupled to the same field strength
  (see the Appendix).}  In particular when the electric and magnetic
charges are equal, $Q=Q_{\rm e}=Q_{\rm m}$, so that the dyonic strings
are self-dual, then the quantisation rule (\ref{dyond6}) becomes 
\be 
Q^{(1)}\,
Q^{(2)} = \pi\, \kappa_6^2\, n\ .\label{dd6quant} 
\ee From this, it follows that the charge of the self-dual string acquires
an absolute scale, given by 
\be 
Q=\kappa_6\, \sqrt{\pi}\, m\ , 
\ee
where $m$ is an integer.  As in the previous cases, the Dirac
quantisation condition arises only when the two self-dual strings have
non-aligned spatial world-volume directions.

     Now let us consider the type IIB theory in $D=10$, where there
exists a self-dual 5-form field strength, which can support a
self-dual 3-brane. The Dirac quantisation condition for the self-dual 3-brane
is once again symmetric, and analogously to the case of the self-dual string in
$D=6$, one finds that the charge
$Q$ satisfies the condition
\be
Q^{(1)}\, Q^{(2)} = \pi\, \kappa_{10}^2\, n\ .\label{d10quant}
\ee
This implies that there is again an absolute scale for the charge
of the self-dual 3-brane, namely
\be
Q= \kappa_{10}\, \sqrt{\pi}\, m\ ,\label{d10abs}
\ee
where $m$ is an integer.

One way to understand the origin of the quantisation condition
(\ref{d10quant}) for self-dual 3-branes is by studying the Dirac
quantisation condition for black holes in $D=4$ that arise from
3-branes wrapping around the internal coordinates.  The 2-form field
strengths $\F2{\a\b}$ in $D=4$, where the usual $i,j\cdots$ indices
have been split as $i=(1,2,\a)$, {\it etc.}, can also be obtained by
dimensional reduction from the self-dual 5-form in the type IIB
theory. In particular, this implies that both electric and magnetic
black holes using a given $\F2{\a\b}$ can be oxidised to give rise to
self-dual 3-branes in $D=10$. Furthermore, in $D=4$ only the electric
and magnetic charges for the {\it same} field strength $\F2{\a\b}$
suffer a Dirac quantisation condition.  As we saw in section
\ref{dirac}, this pair of black holes will oxidise to give a pair of
3-branes in $D=10$ with no overlapping spatial worldvolume
coordinates, in accordance with the discussion in section
\ref{sec:charges} showing that Dirac quantisation conditions arise
only in cases with no spatial orientation coincidences. Thus, the
quantisation condition (\ref{d10quant}) for the self-dual 3-brane
should be obtained as a consequence of the ordinary $D=4$ quantisation
condition for black holes.

In order to make the relation between quantisation conditions in the
various dimensions more precise, one needs to be careful with certain
factors that arise in the dimensional reduction process for self-dual
fields. The dimensional reduction of the $D=10$ self-dual 5-form
yields both a 4-form $H_4$ and a 5-form $H'_5$ in $D=9$; the $D=10$
self-duality condition then yields the $D=9$ condition that the 5-form
is in fact the dual of the 4-form: $H'_5= \wtd H_4$. Upon eliminating
$H'_5$ from the formalism, one obtains a theory involving only a
4-form $H_4$, but this form does not yet have the canonical
normalisation.  In order to correct for its non-standard
normalisation, one needs to rescale it: $H_4^{\rm
  canonical}=\sqrt{2}H_4$. Accordingly, charges defined using
$H_4^{\rm canonical}$ and its further dimensional descendants need to
be scaled up by $\sqrt2$ in order to be compared with the charge scale
set for the $D=10$ self-dual 3-brane by (\ref{d10abs}); for example,
for the $D=9$ electric 2-brane and the dual magnetic 3-brane, both
supported by $H_4^{\rm canonical}$, one obtains the quantisation
condition 
\be 
Q^{9}_{\rm e}\, Q^{9}_{\rm m} = 2\pi\, \kappa_{9}^2\, n\ 
,\label{q9quant} 
\ee 
as expected. Further dimensional reductions of the pure electric and
pure magnetic solutions in $D=4$ proceed as discussed earlier, with
the relations between the charges as given in Table 1. Accordingly,
the $D=10$ quantisation condition (\ref{d10quant}) both implies and is
implied by the standard $D=4$ black-hole Dirac quantisation conditions.

A similar argument to the above, in which a self-dual string in $D=6$
is obtained by oxidation from a dual pair of 2-charge black-holes in
$D=4$, gives the quantisation condition (\ref{dd6quant}).  To see
this, we note that a self-dual string can be first diagonally and then
vertically reduced to an electric black hole in $D=4$.  A second
self-dual string with no spatial orientation coincidences relative to
the first can first be vertically and then diagonally reduced, giving
the magnetic dual black hole in $D=4$ that is dual to the electric one
first obtained.  The Dirac quantisation condition for the
electric/magnetic black holes in $D=4$ then implies the quantisation
rule for non-aligned self-dual strings in $D=6$, after taking into
account a factor of $1/\sqrt{2}$ rescaling analogous to that discussed
above for the $D=10$ 3-brane case.\footnote{The details of the
  dimensional reduction of self-dual strings in $D=6$ have been given
  in Ref.\ \cite{stainless}.}  By taking the electric and magnetic
charges on the black holes in $D=4$ to be independent, one can
similarly derive the Dirac quantisation rule (\ref{dyond6}) for dyonic
strings in $D=6$.

\section{Quantisation conditions, intersections and supersymmetry}
\label{intersections}

     In the previous sections, we have seen that there are several
subtleties that arise in the Dirac quantisation conditions for extended
objects. In particular, the relative orientation of a conjugate pair of
electric and magnetic objects is important for determining whether or
not a quantisation condition between them arises.  Relative
orientations are also important in the discussion of intersecting
$p$-branes.  In fact, the fraction of preserved supersymmetry is
determined by the relative orientation.  This suggests that there is
a close relationship in M-theory or in type II string theory between
the preservation of supersymmetry, intersections and the Dirac quantisation
conditions for extended objects.

     Intersecting $p$-branes can be viewed as the higher-dimensional
oxidations of multi-charge $p$-branes in a lower dimension. We shall
concentrate in particular on intersecting $p$-branes that reduce to multi black
hole solutions in $D=4$.  Since supersymmetry is preserved under dimensional
reduction, it follows that the relationships that we have mentioned above can
equally well be addressed in a four-dimensional framework.  In $D=4$, there
are a total of 28 2-form field strengths.  Thus there can be in total 56
charges, comprised of 28 electric and 28 magnetic charges.  Any pair
of these can be used to construct a 2-charge black hole solution.
However, there are three quite distinct kinds of 2-charge solution
that can arise, which can be distinguished by the fraction of
supersymmetry that they preserve.  Let us suppose that the dilaton
vectors for the two field strengths are $\vec c_1$ and $\vec c_2$.  We
may then define the quantity $\tilde\Delta$ by 
\be 
\tilde\Delta =
\ft14(\epsilon_1\, \vec c_1 + \epsilon_2\, \vec c_2)^2 +1\ ,
\label{delta}
\ee
where $\epsilon$ is $+1$ if the field carries an electric charge, and
$-1$ if it carries a magnetic charge.  The three possible kinds of
2-charge black-hole can then be characterised by $\tilde\Delta$, as given in
Table 2 below \cite{classp}.

\bigskip\bigskip

\centerline{
\begin{tabular}{|c|c|c|c|}\hline
 & Mass & $\begin{array}{c}\hbox{\bog} \cr \hbox{eigenvalues}
                   \end{array}$ & Supersymmetry \\ \hline\hline
$\tilde\Delta=3$ & $m=\sqrt{Q_1^2+Q_2^2}$ & $\mu=m\pm\sqrt{Q_1^2+Q_2^2}$ &
$\ft12$ \\ \hline
$\tilde\Delta=2$ & $m=|Q_1| + |Q_2|$ & $\mu=m\pm |Q_1| \pm |Q_2|$ &
$\ft14$ \\ \hline
$\tilde\Delta=1$ & $m=(Q_1^{2/3}+Q_2^{2/3})^{3/2}$ 
& $\mu=m\pm\sqrt{Q_1^2+Q_2^2}$ &
$0$ \\ \hline
\end{tabular}}
\bigskip

\centerline{Table 2: The three kinds of 2-charge black hole}
\bigskip
\noindent
The $\tilde\Delta=3$ solution is nothing but a transformation of a
single-charge black hole under an $SL(2,\R)$ subgroup of the $E_7$
U-duality group, and so it preserves $\ft12$ of the supersymmetry.
The $\tilde\Delta=2$ solution is a `genuine' 2-charge solution, which
cannot be reduced to a single-charge solution by any duality
transformation.  As usual for such solutions, it preserves $\ft14$ of
the supersymmetry.  The $\tilde\Delta=1$ solution is a dyonic black hole
\cite{gk} where the electric and magnetic charges are carried by the
same field strength.  This solution preserves no supersymmetry.
In all cases the fractions of preserved supersymmetry can be read off
from the eigenvalues of the \bog matrix, which is constructed as the
anticommutator of the eleven-dimensional supercharges \cite{dlps,lpsol}.
Specifically, each of its zero eigenvalues corresponds to an unbroken
component of supersymmetry.  The $\tilde\Delta=3,2$ and 1 solutions are
described in terms of 1, 2 and 0 harmonic functions respectively.

     As we have discussed previously, the Dirac quantisation
conditions in four dimensions arise only between a pair of electric and
magnetic charges that are carried by the same field strength.  As we have seen
above, such a pair of charges arise in the $\tilde\Delta=1$ dyonic black hole
solution. Oxidising the three kinds of 2-charge black holes back to $D=11$,
they all become intersections of M-branes, waves or NUTs.  In particular, a
membrane and a 5-brane can have three possible kinds of intersection,
with three kinds of relative orientation, depending on whether they
originate from $\tilde\Delta=3, 2$ or 1 black holes in four
dimensions. These can be called non-harmonic, harmonic and
non-supersymmetric intersections respectively \cite{classp}. Specifically,
the non-harmonic intersections occur when the membrane and 5-brane share two
common spatial world-volume coordinates; in harmonic intersections they share
one such coordinate; and in the non-supersymmetric intersections there are no
common spatial world-volume coordinates. Thus we see that whenever the
relative orientations of the membrane and 5-brane are such that they 
give supersymmetric intersections, there is no Dirac quantisation condition
between them.

Note that although we have taken the case of a membrane and a 5-brane
as an example, a similar conclusion applies to all the other possible
intersections in $D\le11$, namely: brane intersections that preserve
any degree of supersymmetry are not subject to Dirac quantisation
conditions.  While we shall not give a formal proof of the
correspondence between supersymmetric and Dirac-insensitive
configurations, one could most likely be worked out by exploiting the
relation between the Dirac quantisation conditions and the
quantisation of angular momentum in the presence of the field
corrections to the angular momentum of dual pairs of electric and
magnetic objects, generalising the classic discussion for dyonic point
particles \cite{angmom}.

\section{Dirac quantisation conditions and the M-theory conjecture}
\label{web}

           So far we have discussed the quantisation conditions 
for a variety of different charges.  For some, like those of 
Kaluza-Klein waves
and NUTs, the electric and magnetic charges are separately discretised
in their own right, independent of (but consistent with) 
any Dirac quantisation condition.
In other words, the absolute units of electric and magnetic charge are
separately determined in these cases.  In the case of self-dual
$p$-branes, the absolute unit of charge is also determined, by virtue
of the Dirac quantisation condition and its symmetric form in $D=4k+2$
dimensions.  For more general $p$-branes, however, the charges satisfy
only the ordinary Dirac quantisation condition, and this
fixes only the {\it product} of the electric and magnetic charges 
associated with each field strength, leaving the absolute units for
each of the charges undetermined.  Thus in
$D$-dimensions, the most general minimum-charge solution to the Dirac 
quantisation condition $Q_e
Q_m=2\pi \kappa_{\sst D}^2$ is given by\footnote{Note that in the
following discussion, we shall always present only the minimum
charge units.  The allowed charges are then any integer multiples of
these minimum units.}
\be
Q_{\rm e} = (2\pi)^{\gamma}\, \kappa_{\sst D}^{2(D-n-1)/(D-2)}\ ,\qquad 
Q_{\rm m} = (2\pi)^{1-\gamma}\, \kappa_{\sst D}^{2(n-1)/(D-2)}  
\ .\label{charges2}
\ee
where $\gamma$ is a free parameter. Here we have made use of the fact that
the charges and $\kappa_{\sst D}$ have the following engineering 
dimensions
\be
{[}Q_{\rm e}{]}= L^{D-n-1}\ ,\qquad {[}Q_{\rm m}{]}= L^{n-1}\ ,\qquad
{[}\kappa_{\sst D} {]} = L^{D/2-1}\ .
\ee
The degree of freedom parameterised by $\gamma$ is obviously undesirable
in a theory that is
believed to be fundamental.  For example, in the type IIA string
theory there are three field strengths in all, namely $F_4$, $F_3^{(1)}$
and $F_2^{(1)}$, whose electric and magnetic charges are each subject
to Dirac quantisation conditions.  There are therefore three as-yet 
undetermined parameters $\gamma_n$
associated with the absolute charge scales of the three pairs of 
minimum electric and magnetic charges, namely
\be
Q_{e(n)}^{A}=(2\pi)^{\gamma_n} \kappa_{\sst A}^{(9-n)/4}\ ,\qquad
Q_{m(n)}^{A}=(2\pi)^{1-\gamma_n} \kappa_{\sst A}^{(n-1)/4}
\label{d10Acharges}
\ee
for the field strengths of degree $n=4,3$ and 2, where the $A$
superscripts indicate charges in the type IIA theory.  On the other
hand, string theory is supposed to have only one free parameter,
namely the string tension.  Purely within the string theory itself, it is
hard to see how these extra parameters in the spectrum of states can be
fixed.

   It has been conjectured that the strong-coupling limit of type
II string theory is described by a theory in eleven dimensions, known as
M-theory.  Its low-energy limit is described by the usual
eleven-dimensional supergravity Lagrangian.  The membrane in $D=11$
can be double-dimensionally reduced to give the perturbative NS-NS
string of the type IIA theory in $D=10$.  It can alternatively be
vertically reduced to give an R-R membrane in $D=10$.  Similarly, the
5-brane in $D=11$ can be vertically reduced to an NS-NS 5-brane in
$D=10$, or diagonally reduced to an R-R 4-brane.  The M-theory
conjecture implies that the string and membrane charges in $D=10$ have
scale sizes related by the compactification period $L_1$, since, as we
showed in
section \ref{dirac}, dimensional reduction yields canonically defined charges
related as shown in Table 1. Similarly, the 5-brane and 4-brane
charges in $D=10$
should have related scale sizes.
Furthermore, the D0-brane and D6-brane in type
IIA theory now are interpreted as the dimensional reductions of the 
Kaluza-Klein waves and
NUTs, whose charge discretisations are, as we have seen in section 5, already 
absolutely determined. Thus with the introduction of M-theory the
number of  undetermined
charge parameters in the type IIA string is reduced from three to only
one, namely
\be
Q_{e(4)}^{M}=(2\pi)^{\a} \kappa_{11}^{4/3}\ ,\qquad
Q_{m(4)}^{M}=(2\pi)^{1-\a} \kappa_{11}^{2/3}\ .
\ee
Here the superscript $M$ indicates M-theory charges, and the
numerical subscripts on the charges indicate the degree of the
associated field strength.  The duality
between the type IIA string and M-theory implies that $\kappa_{11}^2 = 
\kappa_{\sst A}^2 L_1$, where $L_1$ is the period of the compactifying
coordinate $z_1$.  Furthermore, we can express the three
parameters $\gamma_{n}$ in (\ref{d10Acharges}) in terms of the
single parameter $\a$, by equating the
type IIA charges to those coming from the dimensional
reduction of M-branes, waves and NUTs in $D=11$.  Thus we have
\be
(2\pi)^{1-\gamma_1} = \Big(\fft{\kappa_{11}^2}{L_1^{9}}\Big)^{-1/8}\ ,\quad
(2\pi)^{\a - \gamma_2} =\Big(\fft{\kappa_{11}^2}{L_1^{9}}\Big)^{1/12}\ ,\quad
(2\pi)^{\gamma_3 -\a} =\Big(\fft{\kappa_{11}^2}{L_1^{9}}\Big)^{1/24}\ .\quad
\ee

       Although M-theory reduces the three free parameters in the spectrum
of the type IIA string to one, there still seems to be no mechanism within the
theory itself for determining the absolute quantised values of the
M-brane charges.  The
situation changes, however, if one also takes into account the
T-duality that relates the type IIA and type IIB strings when they are
compactified on a circle.  In fact this perturbative T-duality enables
us to fix the absolute scales of all the charges.  To see this, recall that
the type IIB theory has the following field content in $D=10$: the NS-NS
sector comprises the metric, the dilaton and a 2-form potential, while
the R-R sector comprises an axion, another 2-form potential and a 4-form
potential whose field strength is self-dual.  Compactification of the
type IIB theory on a circle gives a theory that can
be related to the compactification of the type IIA theory on a circle of
inverse radius.  The relations between the gauge potentials of these
two nine-dimensional theories (including the axions) are summarised in
Table 3.

\vfill\eject

\bigskip\bigskip
\begin{center}
\begin{tabular}{|c|c|c|c|c|c|}\hline
    &\multicolumn{2}{|c|}{IIA} & 
    &\multicolumn{2}{c|}{IIB} \\ \cline{2-6}
    & $D=10$ & $D=9$ &T-duality & $D=9$ & $D=10$ \\ \hline\hline
    & $A_3$ & $A_3$ & $\longleftrightarrow$ & 
                   $A_3$ & ${\bf B}_4$ \\ \cline{3-6}
R-R & &  $A_2^{(1)}$& $\longleftrightarrow$ 
                           & $A_2^{\rm R}$ & $A_2^{\rm R}$
                                               \\ \cline{2-5}
fields& ${\cal A}_1^{(1)}$ & ${\cal A}_1^{(1)}$ & 
                $\longleftrightarrow$ &
        $A_1^{\rm R}$ & \\ \cline{3-6}
   & & ${\cal A}_0^{(12)}$ & $\longleftrightarrow$ 
                            & $\chi$ &$\chi$
                                 \\ \hline\hline
NS-NS & $G_{\mu\nu}$ & $\cA_1^{(2)}$ 
                        & $\longleftrightarrow$ &
        $A_1^{\rm NS}$ & $A_2^{\rm NS}$ \\ \cline{2-5}
fields& $A_2^{(1)}$ & $A_2^{(1)}$ &
               $\longleftrightarrow$ & $A_2^{\rm NS}$ &
                                       \\ \cline{3-6}
      & & $A_1^{(12)}$ & $\longleftrightarrow$ & 
                              $\cA_1$ & $G_{\mu\nu}$
                                       \\ \hline
\end{tabular}
\end{center}

\bigskip\bigskip

\centerline{Table 3: Gauge potentials of type II theories in $D=10$
and $D=9$}
\bigskip\bigskip

    The relation between the dilatonic scalars of the two
nine-dimensional theories is given by
\be
\pmatrix{\phi \cr \varphi}_{IIA} =\pmatrix{\ft34 & -\ft{\sqrt7}{4} \cr
                                           -\ft{\sqrt7}{4} & -\ft34}
\pmatrix{\phi \cr \varphi}_{IIB} \ .\label{dils}
\ee
The dimensional reduction of the ten-dimensional string metric to
$D=9$ is given by
\bea
ds_{\rm str}^2 &=& e^{-\ft12\phi}\, ds_{10}^2 \nn\\
&=& e^{-\ft12\phi}\, (e^{\varphi/(2\sqrt7)}\, ds_9^2 +
e^{-\sqrt7\varphi/2} \, (dz_2 + {\cal A})^2 ) \ ,
\eea
where $ds_{10}^2$ and $ds_9^2$ are the Einstein-frame metrics in
$D=10$ and $D=9$.  The radius of the compactifying circle, measured using
the ten-dimensional string metric, is therefore given by $R=e^{-\ft14
\phi -\sqrt7\varphi /4}$.  It follows from (\ref{dils}) that the radii
$R_{IIA}$ and $R_{IIB}$ of the compactifying circles, measured using
their respective ten-dimensional string metrics, are related by
$R_{IIA}=1/R_{IIB}$.

        Now as we discussed in section 6, there is an absolute charge
quantisation for the
self-dual 3-brane supported by the self-dual 5-form field strength in
type IIB.  On other hand, the charges for the NS-NS and R-R 3-form field
strengths each have an as-yet undetermined scale parameter.  Thus the type IIB
charges in $D=10$ are given by
\bea
Q_{(5)}^{B} &=& \sqrt{\pi}\, \kappa_{\sst B}\ ,\nn\\
Q_{e(3)}^i &=& (2\pi)^{\beta_i}\, \kappa_{\sst B}^{3/2}\ ,\qquad
Q_{m(3)}^i = (2\pi)^{1-\beta_i}\, \kappa_{\sst B}^{1/2}\ .
\eea
where $i=1$ and 2 label the charges of the NS-NS and the R-R 3-form field
strengths respectively.  Note that we have omitted the charge for the
axion $\chi$, which we shall discuss later.  Thus before applying the
T-duality relation, the charges in M-theory and in type IIB have a
total of three parameters, namely $\a$, $\beta_1$ and $\beta_2$.

    The dimensional reductions of M-theory and the type IIB theory to $D=9$
give rise to a total of six field strengths with degrees $\ge 2$ in each
case, as presented in Table 3.  The T-duality between the two theories
implies that the electric charges and the magnetic charges for the
related pairs of IIA and IIB fields, as indicated in Table 3, should
be set equal.
This gives rise to a total of twelve equations of constraint.  However,
the product of the electric and magnetic charges for each field strength is
the same, simply giving rise to a constraint between the Newton constants
of the theories:
\be
\fft{\kappa_{11}^2}{L_1 L_2} = \fft{\kappa_{\sst A}^2}{L_2} =
\fft{\kappa_{\sst B}^2}{L_{\sst B}}
\ .\label{kapparelation}
\ee
(We are denoting by $L_1$ and $L_2$ the periods of the compactifying
coordinates $z_1$ and $z_2$ in the descent from M-theory, and by
$L_{\sst B}$ the period of the compactifying coordinate in the descent from the
type IIB theory.)
This leaves us with six equations still to consider.  The dimensional 
reduction for canonically-defined charges was given in Table 1.  
Without loss of
generality, we shall present the equations relating just the electric charges
of the two theories, following the order presented in Table 3.  Thus
for the R-R sector we have
\bea
\fft{(2\pi)^\a \kappa_{11}^{4/3}}{L_1 L_2} &=&\sqrt{2\pi}\, \kappa_{\sst
B}\ ,\nn\\
\fft{(2\pi)^\a \kappa_{11}^{4/3}}{L_1} &=&
\fft{(2\pi)^{\beta_2} \kappa_{\sst B}^{3/2}}{L_{\sst B}} \ ,
\label{rrcons}\\
\fft{2\pi\kappa_{11}^2}{L_1^2 L_2} &=& 
(2\pi)^{\beta_2} \kappa_{\sst B}^{3/2}\ ,\nn
\eea
and for NS-NS sector we have
\bea
\fft{2\pi \kappa_{11}^2}{L_1 L_2^2} &=& (2\pi)^{\beta_1} 
\kappa_{\sst B}^{3/2}\ ,\nn\\
\fft{(2\pi)^\a \kappa_{11}^{4/3}}{L_2} &=&
\fft{(2\pi)^{\beta_1} \kappa_{\sst B}^{3/2}}{L_{\sst B}} \ ,
\label{nscons}\\
(2\pi)^\a\kappa_{11}^{4/3} &=& 
\fft{2\pi \kappa_{\sst B}^{2}}{L_{\sst B}^2}\ .\nn
\eea
The solution to the equations (\ref{kapparelation}),
(\ref{rrcons}) and (\ref{nscons}) is given by
\bea
&&\kappa_{11}^2 = \ft{1}{2\pi} (L_1 L_2 L_{\sst B})^3\ , 
\quad \kappa_{\sst A}^2 =\ft1{2\pi} L_1^2 (L_2 L_{\sst B})^3\ ,\quad
\kappa_{\sst B}^2 =\ft1{2\pi} (L_1L_2)^2 L_{\sst B}^4\ ,\nn\\
&&\a= -\ft23\ ,\qquad
(2\pi)^{3 - 4\beta_1} = \Big(\fft{L_2}{L_1}\Big)^2\ ,\qquad
(2\pi)^{3 - 4\beta_2} = \Big(\fft{L_1}{L_2}\Big)^2\ .\label{solutions}
\eea
Thus we see that all three of the originally-free charge-scale
parameters $\a$, $\beta_1$ and $\beta_2$ 
are determined in terms of parameters within the theories, namely the
periods of the compactifying coordinates.  (Note that the three
Newton constants are also expressed in terms of the periods.) 
In particular, this has the
consequence that the $2\pi$ factors for all the  
electric and magnetic charges cancel out, and thus all the minimum
charge units  can be expressed purely in terms of products of certain 
powers of the three periods $L_1$, $L_2$ and $L_{\sst B}$.  
For example, in M-theory we have
\be
Q_{e(4)}^{M} = (L_1 L_2 L_{\sst B})^2\ ,\qquad
Q_{m(4)}^{M} = L_1 L_2 L_{\sst B}\ ,
\ee
and in the type IIB string the NS-NS and R-R 3-form charges are given by
\bea
&&Q_{e(3)}^1 = L_2 L_1^2 L_{\sst B}^3\ ,\qquad
Q_{m(3)}^1 =L_2 L_{\sst B}\ ,\nn\\
&&Q_{e(3)}^2 = L_1 L_2^2 L_{\sst B}^3\ ,\qquad
Q_{m(3)}^2 =L_1 L_{\sst B}\ .
\eea
It is easy to work out the analogous expressions for all the charges
in the type IIA and type IIB theories, as functions of the three periods.

        Note that the $L$ periods are neither dynamical quantities nor
moduli, and they  are fixed independently of any specific solutions to the
lower-dimensional equations of motion.  Since they arise in combination with
exponentials of the dilatonic scalars in (\ref{met}), it follows that
their values can be adjusted by field redefinitions in which the
dilatonic scalars are shifted by constants. Since the periods have
dimensions of length, they cannot be absorbed completely into the dilaton
exponentials, but one can however fix their values in any convenient fashion,
without loss of generality.  A convenient choice is to take them all to be
equal, and then it follows from (\ref{solutions}) that  we have
\be
L_i = L_{\sst B} \equiv L =(2\pi\, \kappa_{11}^2)^{1/9}\ . 
\label{lfix}
\ee
This eliminates all the free parameters, with all charge units now 
being expressed purely in terms of the eleven-dimensional
Newton constant $\kappa_{11}$.  Thus in this convention, we find that the
above analysis of the T-duality relation between the type IIA and IIB
theories implies that all the canonical charges are absolutely
determined,  and their minimum units are given by
\be
Q_e=L^{D-n-1}\ ,\qquad Q_m = L^{n-1}\ .\label{qfix}
\ee

    The T-duality relationship for the quantised charges also provides
supporting evidence for the $SL(2,\Z)$ U-duality of the type IIB
theory.  In particular, as we have shown, in the convention of
(\ref{lfix}) the units for the
NS-NS and R-R charges are not independent parameters,
rather they are equal.  In this convention, the $SL(2,\Z)$ group
elements, under which the NS-NS and R-R charges form a doublet, are
integers.   On other hand, the 7-brane in the
type IIB theory (supported by the axion $\chi$) also (independently) 
discretises the continuous $SL(2,\R)$ global symmetry group to
$SL(2,\Z)$, owing to the topological structure of the solution.
Indeed, in the latter case the group elements are also pure integers,
since it is easy to verify that the allowed magnetic charges for the axion
$\chi$ are now purely integral, owing to the T-duality.   This is
necessary for consistency with the conjectured non-perturbative
$SL(2,\Z)$ symmetry of the type IIB theory.

    A similar, but somewhat different, discussion of the relations between the
charges (or tensions) of the D-branes and the M-branes has been given in Refs
\cite{dlm,sch,alw}. This argument also makes use of T-duality to relate all
the D$p$-branes in the type IIA and type IIB theories, starting from M theory
({\it i.e.}\ from $D=11$ supergravity). Specifically, the self-dual D3-brane in
$D=10$ type IIB theory can be dimensionally reduced either diagonally to a
D2-brane or vertically to a D3-brane in $D=9$.  Mapping these over to the type
IIA theory, and then oxidising them back to $D=10$, they become a dual
D2-brane and D4-brane pair.  This give rise to another
condition on the charges, namely on their quotient, in
addition to the Dirac condition on their product.  In fact, using the 
T-duality that relates a D$p$-brane to a
D$(p+1)$-brane, one can deduce from this that all of the R-R
charges are integer multiples of absolutely-determined fundamental units.

      One may next relate the NS-NS and R-R charges using the
$SL(2,\Z)$ duality symmetry of type IIB theory, under which the NS-NS and R-R
2-form potentials form a doublet.   Another way to do this is to consider
M-theory, in which the D2-brane and the D4-brane can be  
oxidised up to $D=11$, where they become a membrane and a 5-brane. 
Consequently, the two M-branes are both related to the type IIB self-dual
D3-brane, from which one acquires an additional constraint
\cite{dlm,sch}.  The M-branes can
also be dimensionally reduced to give rise to an NS-NS string and a
5-brane in $D=10$, which implies that the NS-NS and R-R charges should
be related.

            In this paper we made use of additional considerations,
namely that the charges associated with Kaluza-Klein vectors are
absolutely determined by mechanisms that go beyond Dirac quantisation.
In this case not only the product, but also the absolute units of each
of the electric and magnetic charges, are separately fixed.  This,
together with the absolute quantisation of the self-dual charges,
implies that all of the charges in the theory are absolutely determined,
after applying the T-duality relation.  This is to be expected, since
in a fundamental theory there should be no further free parameters
governing the charge spectrum, other than the coupling constant of the
theory.  It is important to note that M-theory alone is not enough to
fix the charge lattice completely, and hence M-theory and the type IIB
string are both necessary parts of the full theory in which all of the
charges are fixed.

           It is interesting, by way of comparison, to analyse the
situation in the case where M-theory is not introduced.  Then, the
type IIA charge units are parameterised by the three quantities
$\gamma_n$, with $n=2, 3$ and 4 for the charges (\ref{d10Acharges})
associated with the field strengths of degree $n$.  In this case, the
type IIA and type IIB theories have, {\it a priori}, a total of five
parameters characterising the charge lattices.  Applying the T-duality
relation between the type IIA and type IIB theories leaves us with two
free parameters, one in the NS-NS sector and the other in the R-R
sector.  Invoking the $SL(2,Z)$ symmetry of the type IIB theory enables
us relate the NS-NS and R-R charges, which removes one more free
parameter.  Thus we see that, purely from a string perspective, the
charge spectrum of the theory is not uniquely fixed, since one free
parameter remains.  It is interesting to note that M-theory by itself
also has one free parameter that sets the scales in {\it its} charge
lattices.  However, when the duality symmetries relating M-theory, the
type IIA string and the type IIB string are all utilised, the
quantised units of all charges become absolutely determined.

\section{Charge lattices and quantisation conditions}
\label{chargelattice}

     Throughout this paper, we have been considering the quantisation
conditions for canonically defined charges descended from the $D=11$
electric and magnetic charges (\ref{charges}) together with the
wave/NUT and $D=10$ 3-brane charge. These conditions have
the effect of restricting the classically allowed families of $p$-brane
solutions to a discrete set at the quantum level, with the allowed
charges forming a charge lattice. In $D=11$, there are conjugate
electric and magnetic charge lattices, which are effectively
one-dimensional since the quantisation condition (\ref{eqxxxx})
involves only the magnitudes of the electric and magnetic charge forms
(for Dirac-sensitive orientations). We have seen that the unit of the
one-dimensional electric charge lattice (and also the magnetic charge
lattice) can be determined using T-duality between the type IIA and
type IIB theories.

  By the arguments given in the last section, all of the
lower-dimensional charge lattices for branes of differing dimension
$p$ also prove to be integers times a fundamental unit given by
(\ref{qfix}).  It should be emphasised, however, that if one considers
a specific lower-dimensional maximal supergravity in isolation,
without regard to its higher-dimensional origin, then the charge
lattices for non-dual pairs of $p$-branes will not necessarily be
related. This is because the Dirac quantisation condition is invariant
under the global symmetry group of the theory. Consequently, there is
no unique charge-lattice solution to the Dirac quantisation condition
alone; from any given Dirac-allowed lattice, others can be constructed
by the action of the global symmetry group. In particular, there would
seem to be no {\it a priori} reason even why the same Dirac-allowed
lattice must arise at different points in the modulus space of the
scalar fields.  On the other hand, as we showed in the previous
section, if the lower dimensional theory is viewed as a dimensional
reduction from eleven dimensions, then the charge lattice in the lower
dimension is fixed uniquely, and turns out to be independent of the
values of the scalar moduli.  Indeed, the lattices are all pure
integers times a fundamental unit (\ref{qfix}). In such a situation,
where the Dirac-allowed charge lattices are independent of the moduli,
the continuous global symmetry group is broken down to a discrete
subgroup, namely the U-duality group \cite{ht}.  This
modulus-independent charge lattice, which cannot be derived purely
from the Dirac quantisation condition in $D$-dimensional supergravity taken
in isolation, plays a crucial r\^ole in the discretisation of the
classical global symmetry group to the U-duality group
\cite{trombone}.

The reduction of the supergravity symmetry group to a discrete
subgroup can also be seen in the structure of the perturbative
counterterms to supergravity, whether they occur with infinite
coefficients in ordinary attempts at quantising (nonrenormalisable)
supergravities, or with finite coefficients as determined in the
effective supergravity theories obtained from superstring theory. For
example, one may consider the simple case of $D=4$, $N=2$ supergravity
coupled to $N=2$ matter, which gives rise to non-renormalisable
counterterms already at the 1 loop level \cite{dvn}. In these one-loop
counterterms, the Maxwell field occurs through the stress tensor
$T^{\mu\nu}$, which is invariant under duality transformations. This
good duality behaviour is actually in excess of the classical
expectations, since the classical Maxwell action itself is not in fact
duality invariant, but instead transforms by an overall phase. This
homogeneous classical transformation behaviour of the action is
sufficient to yield a duality symmetry of the field equations, but
when the classical action is combined with the 1-loop counterterms,
the homogeneity property of the total action's transformation
behaviour is lost. The only transformations for which the phases of
the classical action and the counterterms `align' are those of the
quantum $Z_2$ duality group, reproducing the conclusions independently
obtained by considering the Dirac quantisation condition. Analogous
arguments, taking into account the indications that the first
non-trivial (`3-loop') extended supergravity counterterms
\cite{counterterms, hst} are similarly invariant \cite{hst} under the
full Cremmer-Julia duality groups, imply that the U-duality
discretisations of these groups may also be viewed as arising from the
differences in transformation behaviour between homogeneously
transforming actions and the invariance of the counterterms. Again,
the phases only align after restriction to the discretised $G(\Z)$
duality groups.

      The existence of an unbroken U-duality group at the quantum level
depends on the `registration' between the units of charge sublattices
for the various $p$-branes that are not directly related by the Dirac
quantisation condition. If there were not such a registration, a
U-duality transformation on one field type, mapping between allowed
points on that field's sublattice, would not at the same time map
between allowed points on another field's sublattice. In this sense,
requiring the persistence of an unbroken U-duality at the quantum
level may also be taken as a logical basis for requiring the different
charge units to be brought into registration with each other.

      We have seen that dimensional reduction not only leads to a
charge lattice in the lower dimension that is consistent with the
Dirac quantisation conditions, but that it furthermore selects a unique
lattice, which is independent of the values of the scalar moduli.
This additional condition can be seen from Table 1, where `incidental'
quantisation relations between the various canonically-defined charge
vectors arise even in cases where these are not directly required by
the $D$-dimensional Dirac quantisation conditions (\ref{ddirac})
between conjugate pairs of electric and magnetic charges. For example,
in $D=10$ there are conjugate electric strings and magnetic 5-branes
supported by the same 3-form field strength, for which a quantisation
condition of the form (\ref{ddirac}) naturally occurs (provided the
orientations do not fall into the measure-zero Dirac-insensitive
set). From the point of view of the parent $D=11$ supergravity theory,
however, the 3-form field strength in $D=10$ is just one of the two
descendants of the 4-form field strength of the $D=11$ theory.  The
other descendant is the 4-form field strength in $D=10$.  This 4-form
field strength supports an electric 2-brane solution in $D=10$, which
is the vertical dimensional reduction \cite{vertical} of the 2-brane
in $D=11$ \cite{ds}. Although this ten-dimensional 2-brane is
supported by a different field strength from the ten-dimensional
5-brane, there is nonetheless an `incidental' quantisation relation
between the 2-brane and the 5-brane charges in $D=10$, since their
lattice units are related as shown in Table 1.

   We see therefore that the $D=11$ Dirac quantisation condition
(\ref{eqxxxx}) is the parent of four $D=10$ relations: two genuine
$D=10$ Dirac quantisation conditions, between the string and the
5-brane supported by the 2-form in $D=10$, and between the 2-brane and
the 4-brane supported by the 3-form in $D=10$; plus the two incidental
quantisation relations discussed above. As we discussed in section
\ref{dimred}, the genuine Dirac quantisation conditions are
characterised by the two possible sets of mixed vertical/diagonal
dimensional reductions of the $D=11$ 2-brane and 5-brane, which
maintain the same supporting field strength for their $D=10$
descendants. The incidental quantisation relations, on the other hand,
are characterised by the use of two vertical or two diagonal
reductions, which give $D=10$ solutions supported by two different
field strengths. As we saw in section \ref{dimred}, configurations of
branes in $D=11$ corresponding to these cases do not yield true Dirac
quantisation conditions. The distinction between the genuine and
incidental cases is clearly seen in the period-dependent factors in
Table 1: only for the genuine $D=10$ Dirac quantisation conditions
between conjugate solutions do the period factors $L_i$ cancel
out. The incidental quantisation conditions involve both T duality and
the `absolute' scale-setting properties of IIB self-dual 3-branes and
D0-brane/$(D-4)$-brane pairs in IIA theory, as we have seen.

   From the standpoint of supergravity theories, the canonical charges,
with their universal charge lattices, are the most natural ones to
consider. For comparison with results in the literature, however, it
is appropriate to relate the above canonical-charge results to those
corresponding to different basic definitions of the charges. A
well-known result in the context of $D=4$ $U(1)$ gauge theories with
symmetry breaking {\it via} a Higgs sector is the dependence of the
charge lattice on the vacuum angle $\theta$ of the vacuum as well as
the standard dependence on the unit $e_0$ of electric charge
\cite{wit78}. The `standard' charge lattice to which these results
pertain makes use charges defined differently from the canonical
charges that we have been using. The canonical charges are obtained by
an $SL(2,\R)$ symmetry transformation (whose supergravity
generalisations are the $G$ supergravity duality symmetries) that
precisely removes the $(e_0,\theta)$ modulus dependence of the
standard charge lattice. If the $(\phi,\chi)$ scalar sector is
described by an $SL(2,\R)$ matrix
\be
V=\pmatrix{e^{-{\phi\over2}} & e^{\phi\over2}\, \chi \cr
                             0 & e^{\phi\over2}}\label{viel}
\ee
belonging to the Borel subgroup of $SL(2,\R)$
\cite{trombone}, then the `physical' charge lattice with the coupling
constants restored is given by $Q_{\rm phys}=V_0^{-1}Q$, where $V_0$ is
the matrix $V$ evaluated for the asymptotic scalar values, {\it i.e.}\
the moduli, with $e_0=e^{\phi_0/2}$, $\theta=\chi_0$. One then finds that 
$Q_{\rm phys}$ scales up as $e_0$ is increased, and that the
dependence of the electric and magnetic components of the charges on
the vacuum angle $\theta$ agrees precisely \cite{sen} with
that of Ref.\ \cite{wit78}.

Note that the lattice of physical charges $Q_{\rm phys}=V_0^{-1}Q$ is
still invariant under a quantum duality group, but that this is no
longer simply the group of integer-valued $G(\Z) = SL(2,\Z)$ matrices that
maps the canonical $Q$ charge lattice into itself. Instead, the
$Q_{\rm phys}$ lattice is mapped into itself by the
conjugated\footnote{Note also that the sense of this conjugation is
  opposite to that for the vacuum stability group $H=SO(2)$ which
  leaves the scalar moduli $(\phi_0,\chi_0)$ invariant; the latter is
  conjugated to $\hat H = V_0HV_0^{-1}$. The opposite cojgugation
  properties of these two groups gives rise to the existence of the
  special `self-dual' point $(\phi_0,\chi_0)=(0,0)$ in modulus space,
  at which point the intersection group $\hat G(\Z)\cap\hat H$ is
  maximal. The action of this maximal group agrees with that of the
  Weyl group of the duality group $G$. This occurrence of the Weyl
  group of $G$ as the maximal $\hat G(\Z)\cap\hat H$ intersection
  persists for the complete list of Cremmer-Julia supergravity
  symmetry groups $G$ \cite{weyl,vertical}.} group of matrices $\hat
G(\Z)=V_0^{-1}G(\Z)V_0$.

\section*{Acknowledgements}

We are grateful to Eugene Cremmer and Bernard Julia for useful
discussions on the Dirac quantisation condition in $D=4k+2$
dimensions. We should also like to thank Mike Duff, Sergio Ferrara,
Pietro Fr\'e, Chris Hull, Ioannis Kouletsis, Adam Ritz, John Schwarz and Paul
Townsend for helpful discussions.  We should all like to thank SISSA;
C.N.P. and K.S.S. should like to thank the ENS, and K.S.S. should like
to thank UCLA, for hospitality during the course of the work.

\section*{Appendix: Dirac quantisation conditions in even dimensions}

     In this appendix, we shall discuss further the Dirac quantisation
conditions for charges carried by field strengths of degree $n=\ft12D$
in even dimensions $D$.  For simplicity, we shall consider a single
such field strength, coupled to gravity.  We shall consider both
Lorentzian spacetime signature and Euclidean spacetime signature.

     Consider a Lagrangian of the form
\be
{\cal L} = e\, R -\ft1{2 n!}\, e\, F^2\ ,
\ee
where $F$ is an $n$-form field strength. The energy-momentum tensor is
given by
\be
T_{\mu\nu}= F_{\mu\sigma_2\cdots\sigma_n}\, 
        F_{\nu}{}^{\sigma_2\cdots\sigma_n} -\ft1{2n}\, F^2\,
g_{\mu\nu}\ .\label{enmom1}
\ee
We now define the Hodge dual of $F$ by
\be
\wtd F_{\mu_1\cdots\mu_n} = \ft1{n!}\, 
\epsilon_{\mu_1\cdots\mu_n}{}^{\nu_1\cdots\nu_n}\,
F_{\nu_1\cdots\nu_n}\ ,\label{dual}
\ee
The dualisations have the following properties,
governed by dimension and signature:
\be
\hbox{
\begin{tabular}{lcc}
 & Lorentzian & Euclidean \\
$D=4k$: & $\wtd{\wtd F}= -F$ &  $\wtd{\wtd F}= F$\\
$D=4k+2$: & $\wtd{\wtd F}= F$ &  $\wtd{\wtd F}=- F$\\ 
\end{tabular}}\label{dualrule}
\ee
It is straightforward to see that the energy-momentum tensor
(\ref{enmom1}) can be rewritten using $\wtd F$ in the form
\bea
\hbox{Lorentzian:} &&
T_{\mu\nu}=\ft12 (F_{\mu\sigma_2\cdots\sigma_n}\, 
        F_{\nu}{}^{\sigma_2\cdots\sigma_n} +
         \wtd F_{\mu\sigma_2\cdots\sigma_n}\, 
        \wtd F_{\nu}{}^{\sigma_2\cdots\sigma_n})\ ,\label{loren}\\
\hbox{Euclidean:} &&
T_{\mu\nu}=\ft12 (F_{\mu\sigma_2\cdots\sigma_n}\, 
        F_{\nu}{}^{\sigma_2\cdots\sigma_n} -
         \wtd F_{\mu\sigma_2\cdots\sigma_n}\, 
        \wtd F_{\nu}{}^{\sigma_2\cdots\sigma_n})\ .\label{eucen}
\eea From these expressions, it is clear that when $D=4k$, there is a
continuous symmetry of the energy-momentum tensor under
transformations
\be
F_{\mu_1\cdots \mu_n} \longrightarrow c\, F_{\mu_1\cdots \mu_n} +
                                    s \wtd F_{\mu_1\cdots \mu_n}\ ,
\ee
where the parameters $c$ and $s$ satisfy $c^2+s^2=1$ in the Lorentzian
case, and $c^2-s^2=1$ in the Euclidean case.  Thus in the Lorentzian
case we may take $c=\cos\theta$, $s=\sin\theta$, while in the
Euclidean case we may instead take $c=\cosh\theta$, $s=\sinh\theta$.
On the other hand, in $D=4k+2$ dimensions there is no such continuous
symmetry of the energy-momentum tensor, in either the Lorentzian or
the Euclidean signature.  Thus we have the following duality symmetries in
the various cases, restricting attention to situations involving a single
$n$-form field strength:
\bea
\hbox{Lorentzian:}&& \cases{D=4k: & $SO(2)$ \cr
                            D=4k+2: & \ \ \ -- \cr}\nn\\
\hbox{Euclidean:}&& \cases{D=4k: & $SO(1,1)$ \cr
                            D=4k+2: & \ \ \ -- \cr}
\eea

     In $D=4k$ dimensions we can derive the
Dirac quantisation condition for dyons from the quantisation condition
for purely electric and magnetic charges, by acting on the latter using
the duality symmetry.  Thus, starting from the result that
(dimensionful) charges  $(e_1,0)$
and $(e_2,g_2)$ satisfy the condition $e_1\, g_2 = 2\pi\kappa^2\, n$, one can
deduce that
\be 
e_1\, g_2 -e_2 \, g_1 =2\pi\kappa^2\, n\label{4kdyonic} 
\ee 
for dyons in $D=4k$ dimensions, regardless of the spacetime signature.
By contrast, one cannot deduce the quantisation condition for dyons in
$D=4k+2$ dimensions by any analogous calculation.  In this case the
electric and magnetic charges of the single field strength $F$ cannot
form a doublet under any duality symmetry.\footnote{The absence of a
  continuous duality symmetry in the case of a single $(2k+1)$-form
  field strength in $D=4k+2$ dimensions has been discussed from the
  point of view of the impossibility of realising such a symmetry as a
  canonical transformation in Ref.\ \cite{dght}.}

Of course an enlarged system with more field strengths can have a
duality symmetry, such as in the case of $D=6$ supergravity, where
there is an $O(5,5)$ symmetry.  Using this $O(5,5)$ symmetry, one can
derive the dyonic quantisation condition
\be
e_1\, g_2 + e_2\, g_1 = 2\pi\kappa^2\, n \label{dyonic7}
\ee
as we showed in section \ref{wavesnuts}.  We saw in section \ref{dyons} that
this result holds for dyons in all dimensions $D=4k+2$, using a
generalisation of the arguments of Ref.\ \cite{schwinger}.  Moreover, in
section \ref{wavesnuts}, we saw that the quantisation condition (\ref{dyonic7})
is consistent with dimensional reduction. In section \ref{wavesnuts}, we have
exploited the fact that the Lorentzian $D=4k+2$ quantisation condition
(\ref{dyonic7}) gives rise to a Dirac quantisation condition
for self-dual charges in $D=4k+2$ Lorentzian spacetimes. Note, however, that
there is no such condition for self-dual charges in Euclidean $D=4k$
dimensional spacetimes, since here the dyonic quantisation condition 
(\ref{4kdyonic}) becomes trivial in this case.

\end{document}